\documentclass[10pt,twocolumn,letterpaper]{article}

\usepackage[pagenumbers]{cvpr} 

\definecolor{cvprblue}{rgb}{0.21,0.49,0.74}
\usepackage[pagebackref,breaklinks,colorlinks,allcolors=cvprblue]{hyperref}
\usepackage{graphicx}
\usepackage{multirow}
\usepackage{tabularx}
\usepackage{booktabs}
\usepackage{lipsum}
\usepackage{array} 
\usepackage{svg}
\newcolumntype{C}{>{\centering\arraybackslash}X}

\def\paperID{0005} 
\def\confName{CVPR}
\def\confYear{2025}

\title{OccludeNeRF: Geometric-aware 3D Scene Inpainting with Collaborative Score Distillation in NeRF}

\author{Jingyu Shi\\
Futurewei Technologies\\
Purdue University\\
{\tt\small shi537@purdue.edu}
\and
Achleshwar Luthra\\
Futurewei Technologies\\
Texas A\&M University\\
{\tt\small achleshwarluthra6@gmail.com}
\and
Jiazhi Li\\
Futurewei Technologies\\
University of Southern California\\
{\tt\small jiazhil@usc.edu}
\and
Xiang Gao\\
Futurewei Technologies\\
Stony Brook University\\
{\tt\small gao2@cs.stonybrook.edu}
\and
Xiyun Song\\
Futurewei Technologies\\
{\tt\small xsong@futurewei.com}
\and
Zongfang Lin\\
Futurewei Technologies\\
{\tt\small zlin1@futurewei.com}
\and
David Gu\\
Stony Brook University\\
{\tt\small gu@cs.stonybrook.edu}
\and
Heather Yu\\
Futurewei Technologies\\
{\tt\small hyu@futurewei.com}
}

\begin{document}
\twocolumn[{%
\renewcommand\twocolumn[1][]{#1}%
\maketitle
\centering
\includegraphics[width=.8\linewidth,page=1,trim={0.5cm 0.5cm 0.5cm 0.5cm},clip]{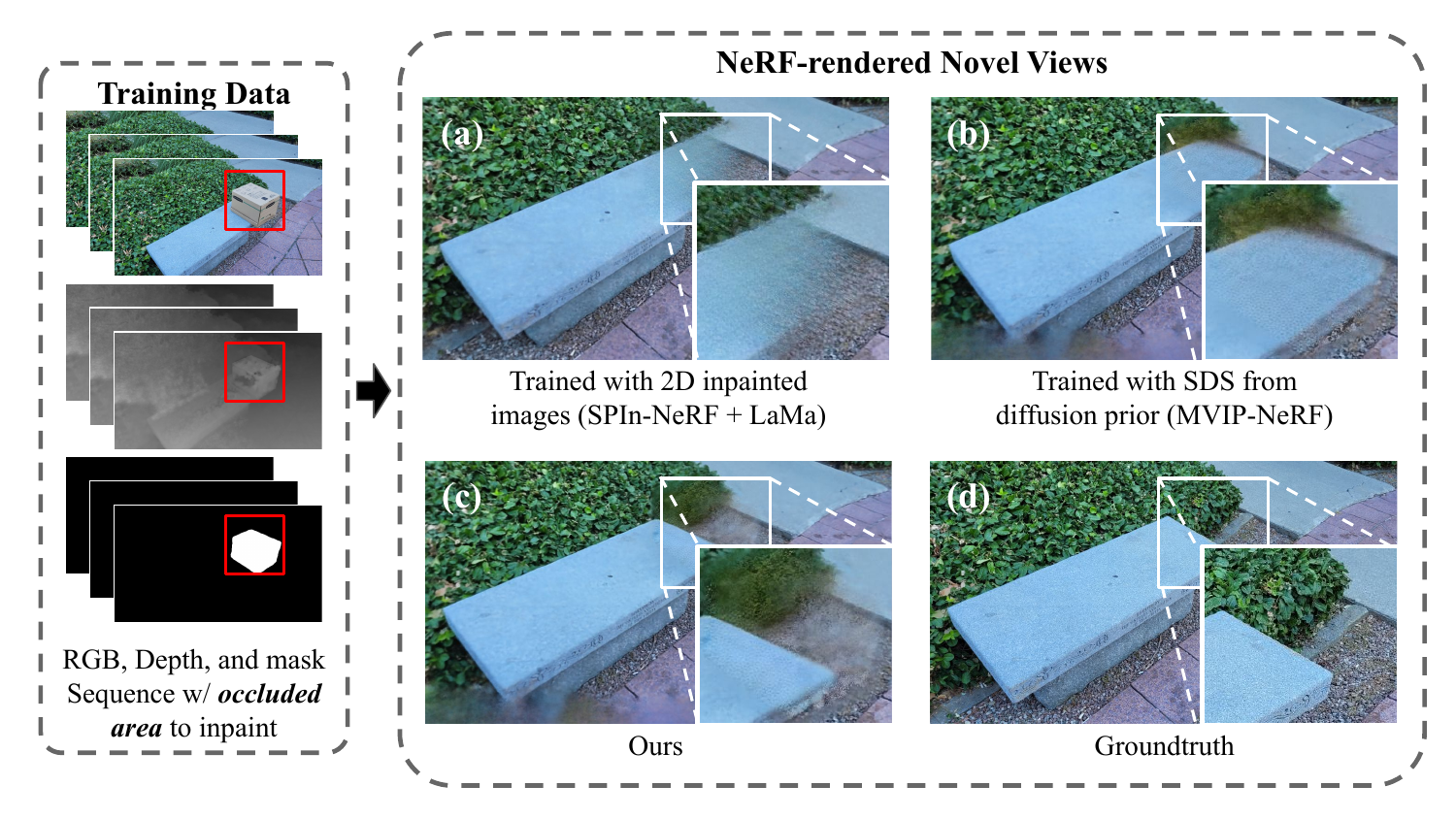}
\captionof{figure}{Given sequences of masked RGB and depth images, prior works (a) and (b) can train a 3D NeRF of the scene with the masked region inpainted.
While showing promising visual fidelity in rendered novel views, the occluded region is not faithfully reconstructed due to limited prior information.
E.g. the rendered bench extends further than that in the groundtruth (d).
Our method (c) incorporates and propagates the limited information to multi-view updates of the NeRF and achieves a more faithful reconstruction of the groundtruth.
\vspace{1em}}
\label{fig:teaser}
}]
\begin{abstract}
With Neural Radiance Fields (NeRFs) arising as a powerful 3D representation, research has investigated its various downstream tasks, including inpainting NeRFs with 2D images.
Despite successful efforts addressing the view consistency and geometry quality, prior methods yet suffer from occlusion in NeRF inpainting tasks, where 2D prior is severely limited in forming a faithful reconstruction of the scene to inpaint.

To address this, we propose a novel approach that enables cross-view information sharing during knowledge distillation from a diffusion model, effectively propagating occluded information across limited views.
Additionally, to align the distillation direction across multiple sampled views, we apply a grid-based denoising strategy and incorporate additional rendered views to enhance cross-view consistency.
To assess our approach's capability of handling occlusion cases, we construct a dataset consisting of challenging scenes with severe occlusion, in addition to existing datasets.
Compared with baseline methods, our method demonstrates better performance in cross-view consistency and faithfulness in reconstruction, while preserving high rendering quality and fidelity.
\end{abstract}
\vspace{-1em}    
\section{Introduction}
\label{sec:intro}

\noindent Neural Radiance Fields (NeRFs)~\cite{mildenhall2021nerf} have emerged as a revolutionary approach to 3D scene representation, showcasing high performance in novel view synthesis.
Recent research has explored diverse NeRF applications across domains, including Augmented and Virtual Reality, game development, and computer-aided design.
Among all, a challenging task is to inpaint a 3D NeRF scene, i.e. to remove an undesired area and complete the area with visually coherent and geometrically plausible content that can be rendered consistently across multiple views.

Inpainting 3D NeRF scenes presents significant challenges.
First, training a NeRF requires 2D images from multiple viewpoints with consistent inpainting to minimize artifacts and enhance visual realism.
Prior approaches address this by focusing on the consistency of 2D inpainted images or incorporating consistent implicit knowledge distillation from diffusion-based 2D inpainting models, such as LaMa~\cite{suvorov2022resolution,mirzaei2023spin,weder2023removing}. 
Alternatively, diffusion models have also gained attention as the 2D inpainter to generate 2d inpainted images~\cite{wang2024innerf360,weber2024nerfiller,lin2025taming}.
In addition to using explicit 2D inpainting, some prior works utilize Score Distillation Sampling (SDS)~\cite{poole2022dreamfusion} to distill generative prior from diffusion models for multiple views~\cite{chen2024mvip,prabhu2023inpaint3d,wang2024innerf360}. 
Despite achieving breakthroughs in rendering quality and multiview consistency, these methods remain challenged by occlusions in 3D NeRF inpainting, where the areas to inpaint are often obscured by objects to be removed, resulting in inconsistent 2D distillation or explicit inpainting over the occluded area across the views.
Such occlusions limit available prior information about the 3D scene, leading to incomplete or unfaithful reconstructions, as shown in ~\cref{fig:teaser}. 

This work addresses the challenge of faithfully reconstructing occluded regions by leveraging information from the limited numbers of occluded views to infer constrained scene priors.
Our approach utilizes multi-view information to guide the inpainting direction, resulting in consistent and faithful reconstructions of the original scenes without compromising rendering quality.

To this end, we present Occlude-NeRF, a novel approach to mitigate the occlusion challenge in 3D NeRF Inpainting while ensuring 3D consistency across multiple views.
Our method uses RGB and Depth images with corresponding binary masks marking inpainted regions as inputs.
We first train a NeRF on the masked images to reconstruct the background.
Meanwhile, to inpaint the masked regions, we followed MVIP-NeRF~\cite{chen2024mvip}, using an SDS training scheme to obtain inpainting guidance from an off-the-shelf diffusion model~\cite{rombach2021highresolution}.
To incorporate information from partially unoccluded views, we apply Collaborative Score Distillation Sampling (CDS)~\cite{kim2023collaborative}, which smooths the gradient update with information from other views and propagates the guidance among the views.
To maximize information sharing among the views, we design a reference-view paradigm during training.
We render two sets of views, with one used for loss back-propagation and the other only leveraged as the reference with no gradient computation. 
To further ensure consistency among multiple views during one distillation step, we applied a grid-denoising pattern in our noise prediction step, inspired by similar findings of prior works~\cite{weber2024nerfiller,bar2022visual}.
By comparison with baseline methods and ablations of our features, we demonstrate how our method handles the occlusion challenge in 3D NeRF inpainting tasks. 
We propose a novel approach to seamlessly and faithfully inpainting severely occluded 3D scenes.
Our code and dataset will be publicly available on GitHub.
Our contributions are listed as follows:
\begin{itemize}
    \item A modified CDS approach with multi-view information sharing in 3D NeRF inpainting task based on diffusion models, to mitigate the occlusion in inpainting areas.
    \item A grid-denoising pattern during score distillation to visually prompt the diffusion model to denoise towards consistent inpainting of distinct viewpoints.
    \item A reference-view training paradigm to increase the cross-view information sharing during NeRF training.
\end{itemize}
To assess the efficacy and performance of our method, in addition to existing datasets~\cite{mirzaei2023spin, mildenhall2019llff}, we construct
\begin{itemize}
    \item A novel and challenging dataset for 3D NeRF inpainting, featuring scenes where the regions to inpaint suffer from occlusion and prior information for inpainting is limited.
\end{itemize}
\section{Related Work}
\label{sec:rw}

\subsection{NeRF Inpainting}

NeRF, or Neural Radiance Fields~\cite{mildenhall2021nerf}, have emerged as powerful representations for synthesizing novel views of complex 3D scenes with high fidelity.
One potential use case of NeRFs is editing or inpainting a scene~\cite{yuan2022nerf,yang2021learning}, which involves filling in or reconstructing missing or corrupted regions to align seamlessly with the context.

NeRF editing can be done by adjusting the color and shape codes~\cite{liu2021editing} supervised by priors from other models~\cite{wang2022clip,mirzaei2022laterf,haque2023instruct}.
Inpainting, however, requires more than straightforward scene editing; it necessitates that inpainted regions visually and geometrically integrate with the original scene.
To tackle these challenges, previous methods have emphasized generating consistent 2D inpainted images~\cite{suvorov2022resolution,cao2021learning,efros1999texture,yu2018generative,nazeri2019edgeconnect} to guide NeRF optimization or training.
Prior methods such as NeRF-In~\cite{liu2022nerf}, SPIn-NeRF~\cite{mirzaei2023spin}, Liu et. al. ~\cite{liu20243d} and Remove-NeRF~\cite{weder2023removing} approach this problem by inpainting 2D images with a 2D inpainter~\cite{cao2021learning,rombach2021highresolution} and constrain the NeRF training with both inpainted 2D images and the 3D consistency among them.

The advent of diffusion models has significantly advanced 2D inpainting capabilities.
Later methods adopt diffusion models as their 2D inpainters while incorporating 3D constraints into the inpainting process, such as visual prompting (NeRFiller~\cite{weber2024nerfiller}), fine-tuning and adversarial training (MALD-NeRF~\cite{lin2025taming}), and 3D self-attention among views (CAT3D~\cite{gao2024cat3d}).

Despite these advancements, methods that rely solely on 2D inpainted images face challenges in achieving complete geometry and multi-view consistency due to variations in 2D inpainting.
MVIP-NeRF~\cite{gao2024cat3d} addresses this limitation by incorporating SDS loss as a multi-view rendering loss, guiding NeRF training with both visual and geometric cues.
This approach has also been utilized in object-level editing ~\cite{zhou2023repaint} and scene-level inpainting~\cite{prabhu2023inpaint3d}.

However, existing methods often struggle when severe occlusion is present in the scene and 2D priors provide limited information about the inpainted regions, complicating faithful scene reconstruction.
Methods like NeRF-W~\cite{martin2021nerf} and Ha-NeRF~\cite{chen2022hallucinated} can remove transient objects from NeRF that occlude the area of interest but are less effective for static objects. 
Zhu et. al.~\cite{zhu2023occlusion} proposed a method to remove the occluding static object from the scene.
However, their method is constrained by the presumption that the occluding object is closer than the background in the scene, while our Occlude-NeRF presents a generalized solution with no such constraints.

Occlude-NeRF builds on the use of SDS loss in NeRF inpainting, focusing specifically on addressing the challenges posed by occlusion.
Specifically, we aim to inpaint and reconstruct the 3D scenes faithfully to the original scenes by enabling information sharing and aligning the distillation direction in SDS.

\subsection{SDS with Diffusion Models}
Diffusion models~\cite{dhariwal2021diffusion,song2020denoising,song2020score,ho2020denoising} refine samples by progressive denoising from noise using a learned denoising process to approximate the target data distribution.
With their ability to model complex data distributions starting from simple ones, such as Gaussian distributions, diffusion models have become the de facto state-of-the-art for image generation.

Beyond sampling in the image space, diffusion models have also been applied to 3D parameter space sampling.
DreamFusion~\cite{poole2022dreamfusion} first introduced the framework of Score Distillation Sampling (SDS), which optimizes parameterized models, such as differentiable image generators like NeRF or 3D Gaussian Splatting (3DGS)~\cite{kerbl20233d}, by distilling rich 2D priors from a pre-trained diffusion model to guide the image generator.
SDS has liberated research in 3D vision from the constraints of expensive 3D training data and inspired numerous variants aimed at tackling different 3D challenges~\cite{hertz2023delta,zou2024sparse3d,kim2023collaborative,zhou2023sparsefusion,wang2024prolificdreamer}, such as enhancing gradient clarity during sampling~\cite{hertz2023delta}, ensuring multi-view consistency~\cite{zou2024sparse3d}.
Among all, Collaborative Score Distillation (CDS)~\cite{kim2023collaborative} facilitates consistent visual synthesis by constraining a smooth vector function within a Reproducing Kernel Hilbert Space (RKHS) while approximating a target distribution through Stein Variational Gradient Descent~\cite{liu2016stein}.
While CDS has primarily been applied to panorama images, video, and 3D scene generation, our work extends its capability for 3D scene inpainting tasks by deriving a parameter update paradigm that enhances multi-view information sharing.

\section{Preliminary}
\label{sec:prelim}

\subsection{Neural Radiance Fields (NeRF)}

Neural Radiance Fields (NeRF) provide a continuous and differentiable method for representing 3D scenes using a neural network.
NeRF parameterizes the scene as a volumetric field that maps a 3D position and a viewing direction to color and density values.
Formally, NeRF can be represented as a function:
\(F_{\theta}(\mathbf{x}, \mathbf{d}) = (\mathbf{c}, \sigma)\),
where \(\mathbf{x} \in \mathbb{R}^3\) denotes a point in 3D space, \(\mathbf{d} \in \mathbb{R}^2\) represents the viewing direction, \(\mathbf{c} \in \mathbb{R}^3\) is the RGB color at the queried point, \(\sigma \in \mathbb{R}^+\) is the volume density, \(\theta\) indicates the learnable parameters of the network.

NeRF employs volume rendering to synthesize images from this 3D representation. The color observed along a camera ray \(\mathbf{r}(t) = \mathbf{o} + t\mathbf{d}\), where \(\mathbf{o}\) is the camera origin and \(\mathbf{d}\) is the direction, is computed as:\(
C(\mathbf{r}) = \int_{t_n}^{t_f} T(t) \sigma(\mathbf{r}(t)) \mathbf{c}(\mathbf{r}(t),\mathbf{d}) \, dt,\)
where \(t_n\) and \(t_f\) denote the near and far bounds along the ray, \(T(t) = \exp\left(-\int_{t_n}^{t} \sigma(\mathbf{r}(s)) \, ds\right)\) represents the accumulated transmittance from \(t_n\) up to point \(t\), describing the probability that the ray has not been occluded.






\begin{figure*}[htp]
  \centering
  \includegraphics[trim={1.2cm .7cm 1.5cm 1cm},clip,width=\linewidth,page=3]{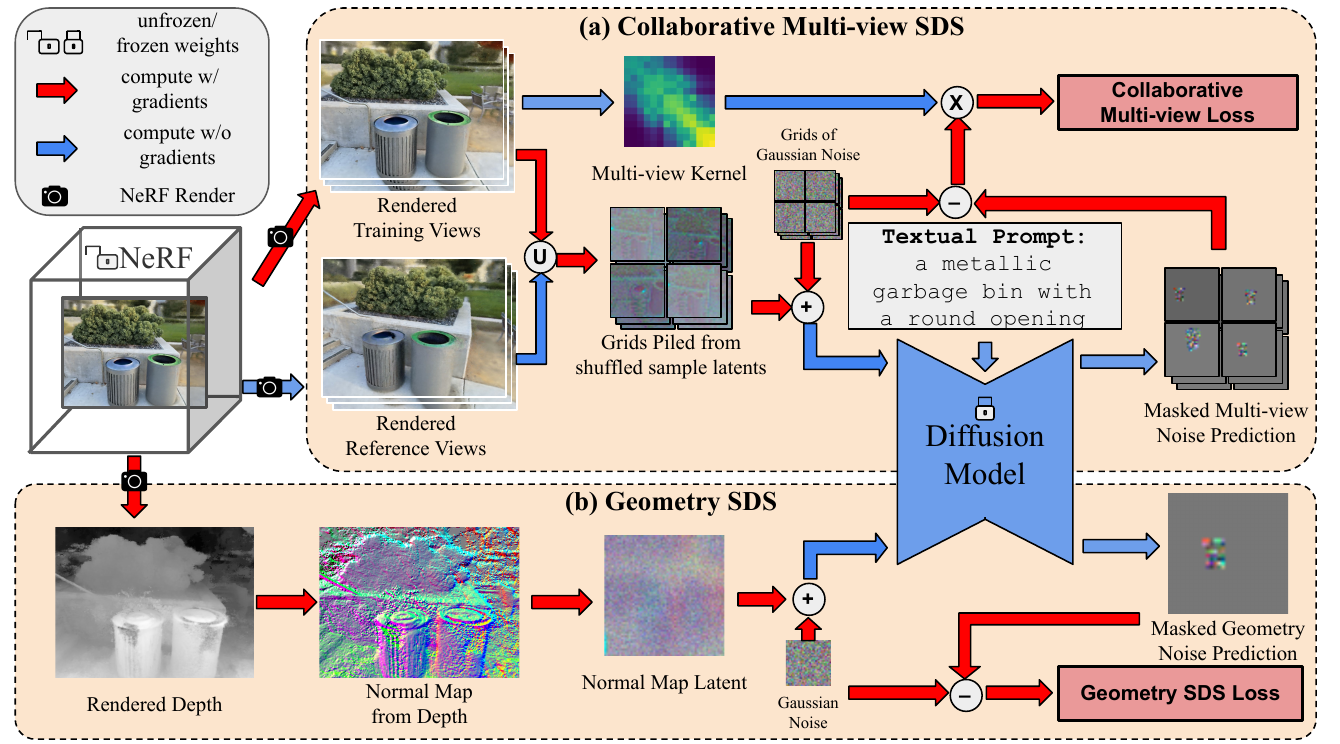}
   \caption{The overall workflow.
   At each iteration, our method takes masked RGB and depth images as input and back-propagates the pixel-wise loss in the unmasked RGB and depth to reconstruct the background (\cref{eq:l_bg}).
   For the masked regions, we render a set of training views and a set of reference views, respectively.
   For the reference views, the gradients are disabled.
   We randomly sample from the union of the two sets and encode grids of latents, which are then added to a grid of Gaussian noise.
   We pass the grids to the diffusion model conditioned on a textual prompt describing the scene and obtain a masked prediction of the noise.
   We then compute a collaborative multi-view loss (\cref{eq:cds_gp}) with a multi-view kernel computed from the training set, assessing how much information to share among the training views.
   We apply a similar geometry SDS loss as in ~\cite{chen2024mvip} (\cref{eq:l_geo}).
   Note that in addition to the masked loss in the figure, we compute the pixel-wise losses in unmasked RGB and depth renderings as well.
   All losses are backpropagated to optimize the NeRF.}
   \label{fig:method}
   \vspace{-1em}
\end{figure*}


\subsection{Score Distillation Sampling}
SDS is an alternative sample generation method proposed by Poole et al.~\cite{poole2022dreamfusion}.
By distilling the knowledge from a 2D text-to-image model, usually a pre-trained diffusion model, SDS optimizes a differentiable image generator (e.g. NeRF or 3DGS) towards a set of 3D parameters that renders high-fidelity images.

Let \(\mathbf{x}=g(\theta)\) be an image rendered by a differentiable generator \(g\) parameterized by \(\theta\).
SDS minimizes the density distillation loss~\cite{oord2018parallel}, which is the KL divergence between the posterior of \(\mathbf{x}\) and the text-conditional density \(p^{\omega}_{\phi}\):
\vspace{-.25em}
\begin{equation}
    L(\theta;\mathbf{x})=\mathbb{E}_{t,\mathbf{\epsilon}}[\alpha_t/\sigma_tD_{KL}(q(\mathbf{x}_t|\mathbf{x})||p^{\omega}_{\phi}(\mathbf{x}_t;y,t))]
    \label{eq:distillationloss}
\vspace{-.25em}
\end{equation}
where \(t\) is the denoising timestep and \(y\) is the embedded textual prompt.
\(\alpha_t\) is the scale and \(\sigma_t\) is the noise variance at \(t\), together defining the noise scheduling.
To update \(\theta\), SDS computes the gradient of the loss by:
\vspace{-.25em}
\begin{equation}
    \nabla_\theta L(\theta;\mathbf{x})=\mathbb{E}_{t,\mathbf{\epsilon}}[w(t)(\mathbf{\epsilon}^{\omega}_{\phi}(\mathbf{x}_t;y,t)-\mathbf{\epsilon})]\frac{\partial\mathbf{x}}{\partial\theta})
\vspace{-.25em}
\end{equation}

where \(w(t)\) is a weighting function.
Derived from SDS, CDS~\cite{kim2023collaborative} aims to update a set of parameters \(\{\theta_i\}^N_{i=1}\) that parameterize the image generator \(g\) for the images \(\mathbf{x}^{(i)}=g(\theta_i)\).
CDS solves the minimization of distillation loss in ~\autoref{eq:distillationloss} by using Stein Variational Gradient Descent (SVGD)~\cite{liu2016stein} in order to update each \(\theta_i\) synchronously within the set \(\{\theta_i\}^N_{i=1}\):
\vspace{-.25em}
\begin{align}
    \nabla_{\theta_i} L(\theta_i;\mathbf{x})&=\frac{w(t)}{N}\sum^N_{j=1}(k(\mathbf{x}^{(j)}_t,\mathbf{x}^{(i)}_t)(\mathbf{\epsilon}^{\omega}_{\phi}(\mathbf{x}^{(i)}_t;y,t)-\mathbf{\epsilon}) \notag \\
    &+\nabla_{\mathbf{x}^{(j)}_t}k(\mathbf{x}^{(j)}_t,\mathbf{x}^{(i)}_t))\frac{\partial\mathbf{x}^{(i)}}{\partial\theta_i})
    \label{eq:cds}
\vspace{-.25em}
\end{align}

where \(k(,):\mathbb{R}^D \times \mathbb{R}^D \rightarrow \mathbb{R}^+\) is a positive definite kernel corresponding to a RKHS.

\section{Methodology}
\label{sec:method}

In this section, we present our proposed method for incorporating multi-view information to address the challenge of occlusion in 3D NeRF inpainting.

Our pipeline is illustrated in ~\cref{fig:method}.
Given a set of RGB images of a scene, corresponding depth images, camera poses, and masks specifying regions to inpaint, our approach trains a NeRF representation of the scene.
The objective is to ensure that the trained NeRF can render novel views with the masked regions consistently inpainted in 3D space. 
For the unmasked background, we train the NeRF with pixel-wise color and depth reconstruction loss:
\vspace{-.25em}
\begin{equation}
    L_{bg} = \lambda_1||\hat{\mathbf{x}}^{(i)}-\Bar{\mathbf{x}}^{(i)}||_2^2 + \lambda_2||\hat{\mathbf{x}}^{(i)}_d-\Bar{\mathbf{x}}^{(i)}_d||_2^2
    \label{eq:l_bg}
\vspace{-.25em}
\end{equation}
where \(\Bar{\mathbf{x}}^{(i)}\) and \(\Bar{\mathbf{x}}^{(i)}_d\) are the masked groundtruth RGB and depth map, and \(\hat{\mathbf{x}}^{(i)}\) and \(\hat{\mathbf{x}}^{(i)}_d\) are the rendered RGB and depth map, with \(\lambda_1,\lambda_2\) being the corresponding weights.

For the masked area, we apply our proposed method to distill RGB prior from a diffusion model to address the occlusion problem in the color space (\cref{eq:cds_gp}).
For geometry supervision of the NeRFs, we perform vanilla SDS for normal map prior, following MVIP-NeRF~\cite{chen2024mvip}:
\vspace{-.25em}
\begin{equation}
    \nabla_\theta L_{geo}(\theta;\mathbf{n}) = w(t)(\mathbf{\epsilon}^{\omega}_{\phi}(\mathbf{z}_t;y,t)-\mathbf{\epsilon})\frac{\partial\mathbf{z}}{\partial\mathbf{n}}\frac{\partial\mathbf{n}}{\partial\theta}
    \label{eq:l_geo}
\vspace{-.25em}
\end{equation}
where \(\mathbf{n}\) is the normal map computed from the rendered depth map. Note that both the geometry and the collaborative losses and noise predictions are computed only within the masked regions.
We also explore guiding geometry with collaborative losses and report the less-satisfactory results in Supplementary~\cref{sec:supp_geocds}.
In the following subsections, we go through the design of our collaborative multi-view loss.
Specifically, we employ a modified version of CDS to collectively update the NeRF parameters using information shared across a subset of views (\cref{subsec:cds}).
We further introduce reference views for collaborative knowledge distillation over more samples (\cref{subsec:ref}).
Finally, we implement a grid-based denoising strategy to enhance cross-view consistency in distillation (\cref{subsec:gp}) and fine-tune the inpainting diffusion model for each scene to ensure visually consistent priors (\cref{subsec:ft}).
\vspace{-.5em}
\begin{figure}[htp]
  \centering
  \includegraphics[trim={1cm 0cm 1cm 0cm},clip,width=\linewidth,page=4]{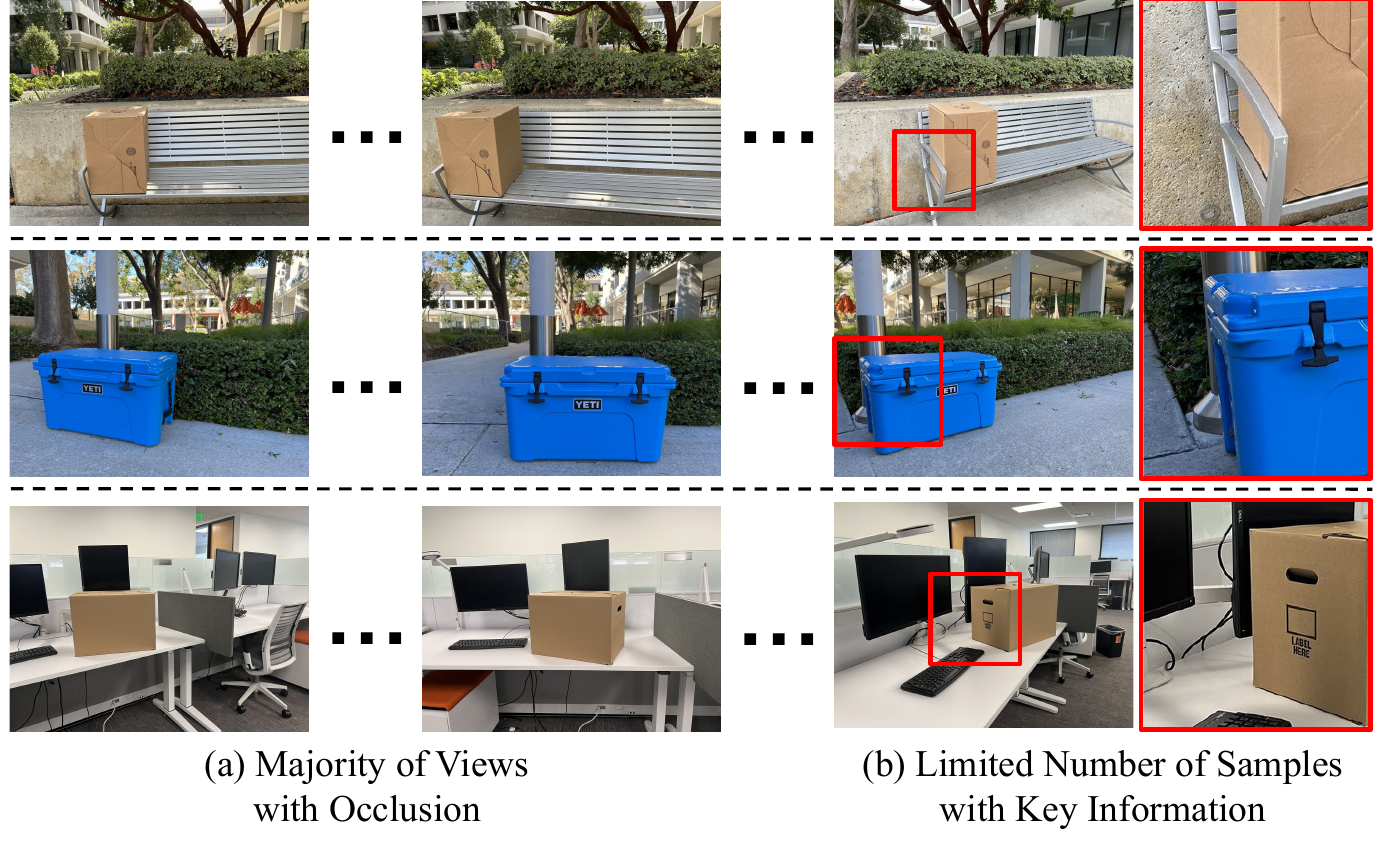}
   \caption{Illustration of the occlusion challenge in NeRF inpainting.
   In the training data, most views are occluded and only a few views have key information for reconstructing the occluded area.
   E.g., the armrest, the pole's shape, and the monitor's edge.}
   \label{fig:challenge}
\end{figure}
\vspace{-.5em}

\subsection{Multi-view CDS}
\label{subsec:cds}
We have witnessed how prior methods fail when the training data has limited information about the area to inpaint due to occlusion.
The core of the challenge, as shown in ~\cref{fig:challenge}, is to extract and propagate faithful information from the limited given views to the parameter updating throughout the NeRF training process.

To tackle this challenge, we aim to train the NeRF progressively with its parameters updated with consideration of multiple randomly selected training views, so that the key information can be extracted and propagated.
To do this, we apply a modified version of CDS (\cref{eq:cds}), where we adopt a radial basis function (RBF) as the kernel:
\vspace{-.25em}
\begin{equation}
    k(\mathbf{x},\mathbf{x'}) = \exp(-\frac{1}{h}||\mathbf{x}-\mathbf{x'}||^2_2)
\vspace{-.25em}
\end{equation}
and denoise multiple views rendered from the NeRF to calculate the distillation loss, starting from:
\vspace{-.25em}
\begin{align}
    &\nabla_{\theta} L(\theta;\mathbf{x})=\frac{w(t)}{N}\sum^N_{i=1}\sum^N_{j=1}(\nabla_{\mathbf{z}^{(j)}_t}k(\mathbf{z}^{(j)}_t,\mathbf{z}^{(i)}_t) \notag \\
    &+k(\mathbf{z}^{(j)}_t,\mathbf{z}^{(i)}_t)(\mathbf{\epsilon}^{\omega}_{\phi}(\mathbf{z}^{(i)}_t;y,\mathbf{m}^{(i)},t)-\mathbf{\epsilon}))\frac{\partial\mathbf{z}^{(i)}}{\partial\mathbf{x}^{(i)}}\frac{\partial\mathbf{x}^{(i)}}{\partial\theta}
    \label{eq:our_cds}
\vspace{-.25em}
\end{align}
where \(\mathbf{z}^{(i)}\) is the encoded latent of \(\mathbf{x}^{(i)}\) by a VAE~\cite{kingma2013auto}.
\(\mathbf{\epsilon}\) is the scheduled noise and \(\mathbf{\epsilon}^{\omega}_{\phi}\) is the predicted noise given
the noised latent \(\mathbf{z}^{(i)}_t\) at \(t\).
\(\mathbf{m}^{i}\) is the concatenation of the masks and unmasked image latent corresponding to \(\mathbf{x}^{(i)}\).
Meanwhile, \(\mathbf{x}^{(i)}\) and \(\mathbf{x}^{(j)}\) are from the same set of rendered views during each update, where \(\mathbf{x}^{(i)}\) is the view to back-propagate from and \(\mathbf{x}^{(j)}\)s are the other views rendered at current iteration. 
We discuss how the first and the second terms spread out the influence of each view and prevent the updates from collapsing into a single mode of target distribution in Supplementary~\cref{sec:supp_cds}.

\subsection{Grid-based Denoising}
\label{subsec:gp}
\begin{figure}[htp]
  \centering
  \includegraphics[trim={1cm 0.8cm 1cm 0.8cm},clip,width=\linewidth,page=5]{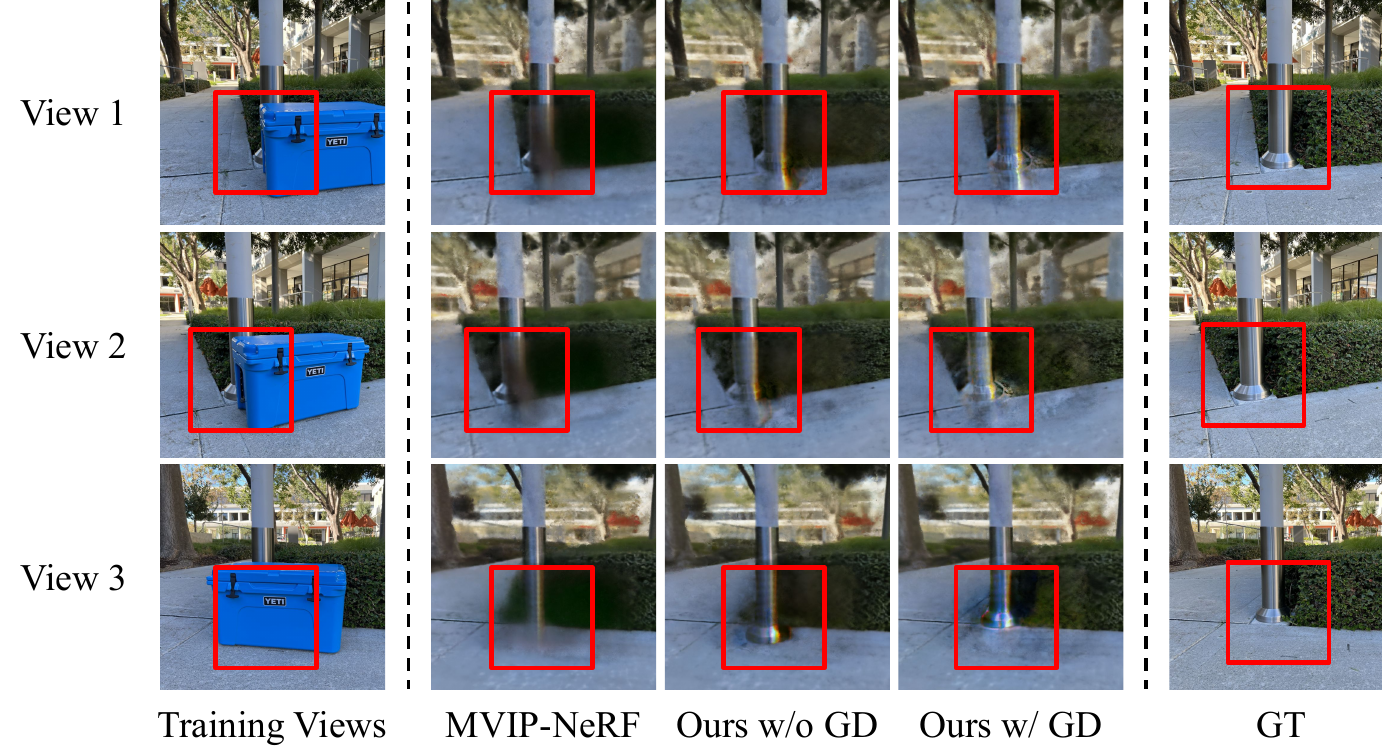}
   \caption{Illustration of the effect of normal denoising and grid-based denoising in distillation. In MVIP-NeRF, vanilla SDS denoising cannot form a consistent denoising direction for the distillation, resulting in blurriness and artifacts.
   With Grid-based Denoising in our method, we observe correct base locations of the pole, reduced artifacts, and clearer inpainting from multiple views.}
   \label{fig:mvipfail}
\end{figure}

As many prior works have pointed out, distilling or training from individually inpainted 2D images results in inconsistent 2D appearance and artifacts in the 3D representations, usually due to the texture shift~\cite{lin2025taming} in the high-frequency area or slight differences in the condition (the background to inpaint)~\cite{chen2024mvip,weber2024nerfiller}. 
Previous SDS method~\cite{chen2024mvip} updates multiple views together with a sum over the distillation loss at each update, which overlooks the directional difference among the noise prediction of each view and still results in blurriness and artifacts, as shown in ~\cref{fig:mvipfail}.

On the contrary, we incorporate the idea of visual prompting that has demonstrated good performance in 2D image space~\cite{weber2024nerfiller,bar2022visual} with our multi-view CDS.
Specifically, at each update, instead of denoising one rendered image at a time, we randomly select multiple rendered training views and pile them into a grid of images.
This grid of images is denoised as a single input, resulting in a corresponding grid of noise prediction, which is ungridded into corresponding noise predictions of each image.

Formally, we incorporate this method into ~\cref{eq:our_cds}:
\vspace{-.25em}
\begin{align}
    &\nabla_{\theta} L(\theta;\mathbf{x})=\frac{w(t)}{N}\sum^N_{i=1}\sum^N_{j=1}(\nabla_{\mathbf{z}^{(j)}_t}k(\mathbf{z}^{(j)}_t,\mathbf{z}^{(i)}_t) + k(\mathbf{z}^{(j)}_t,\mathbf{z}^{(i)}_t)\notag \\
    &(\mathbf{\hat{\epsilon}}^{\omega}_{\phi}(\mathbf{z}^{(i)}_t;\{ \mathbf{z}^{(s)}_t \}_{s=1}^S,y,\mathbf{m}^{(\{s\}+i)}t)-\mathbf{\epsilon}))\frac{\partial\mathbf{z}^{(i)}}{\partial\mathbf{x}^{(i)}}\frac{\partial\mathbf{x}^{(i)}}{\partial\theta}
    \label{eq:cds_gp}
\vspace{-.25em}
\end{align}

where \(\{ \mathbf{z}^{(s)}_t \}_{s=1}^S\) is a subset from the current rendered views randomly selected to form a grid with \(\mathbf{z}^{(i)}\) and \(\mathbf{m}^{(\{s\}+i)}\) is the set of concatenations of corresponding masks and masked image latent.
\(\mathbf{\hat{\epsilon}}^{\omega}_{\phi}\) is the proposed grid-based noise prediction, which is sequentially composed of, (1) a grid operation, (2) a regular noise prediction, and (3) an ungrid operation, to obtain noise predictions corresponding to the input views.
For \(N\) rendered views, we shuffle them, perform grid-based denoising for \(M\) times, and take the average over the \(M\) noise predictions for each view.
Let \(G\) and \(G^{-1}\) be the grid and ungrid operation, respectively:
\vspace{-.25em}
\begin{align}
    G^{-1}(\mathbf{\epsilon}^{\omega}_{\phi}(G(\mathbf{z}^{(i)}_t,&\{\mathbf{z}^{(s)}_t \}_{s=1}^S);y,G(\mathbf{m}^{(\{k\}+i)}),t)) \notag \\
    &=\{\Bar{\mathbf{\epsilon}}^{(i)},\{\Bar{\mathbf{\epsilon}}^{(s)}\}_{s=1}^S\}
\vspace{-.25em}
\end{align}
where \(\{\Bar{\mathbf{\epsilon}}^{(i)},\{\Bar{\mathbf{\epsilon}}^{(s)}\}_{s=1}^S\}\) are the predicted noises of \(\mathbf{z}^{(i)}_t\) and \(\{\mathbf{z}^{(s)}_t \}_{s=1}^S\) respectively.
In short, grid-based denoising treats a grid of images as one, denoises them together, and predicts the noises correspondingly, resulting in consistent denoising directions within the set, as shown in ~\cref{fig:mvipfail}.

To this end, we obtain the grid-based noise prediction w.r.t. the \(i^{th}\) rendered view:
\vspace{-.25em}
\begin{align}
    \mathbf{\hat{\epsilon}}^{\omega}_{\phi}(\mathbf{z}^{(i)}_t;\{\mathbf{z}^{(s)}_t\},y,\mathbf{m}^{(\{s\}+i)}t) = \Bar{\mathbf{\epsilon}}^{(i)}
\vspace{-.25em}
\end{align}

\vspace{-.5em}
\begin{figure}[htp]
  \centering
  \includegraphics[trim={0cm 0cm 0cm .5cm},clip,width=\linewidth,page=6]{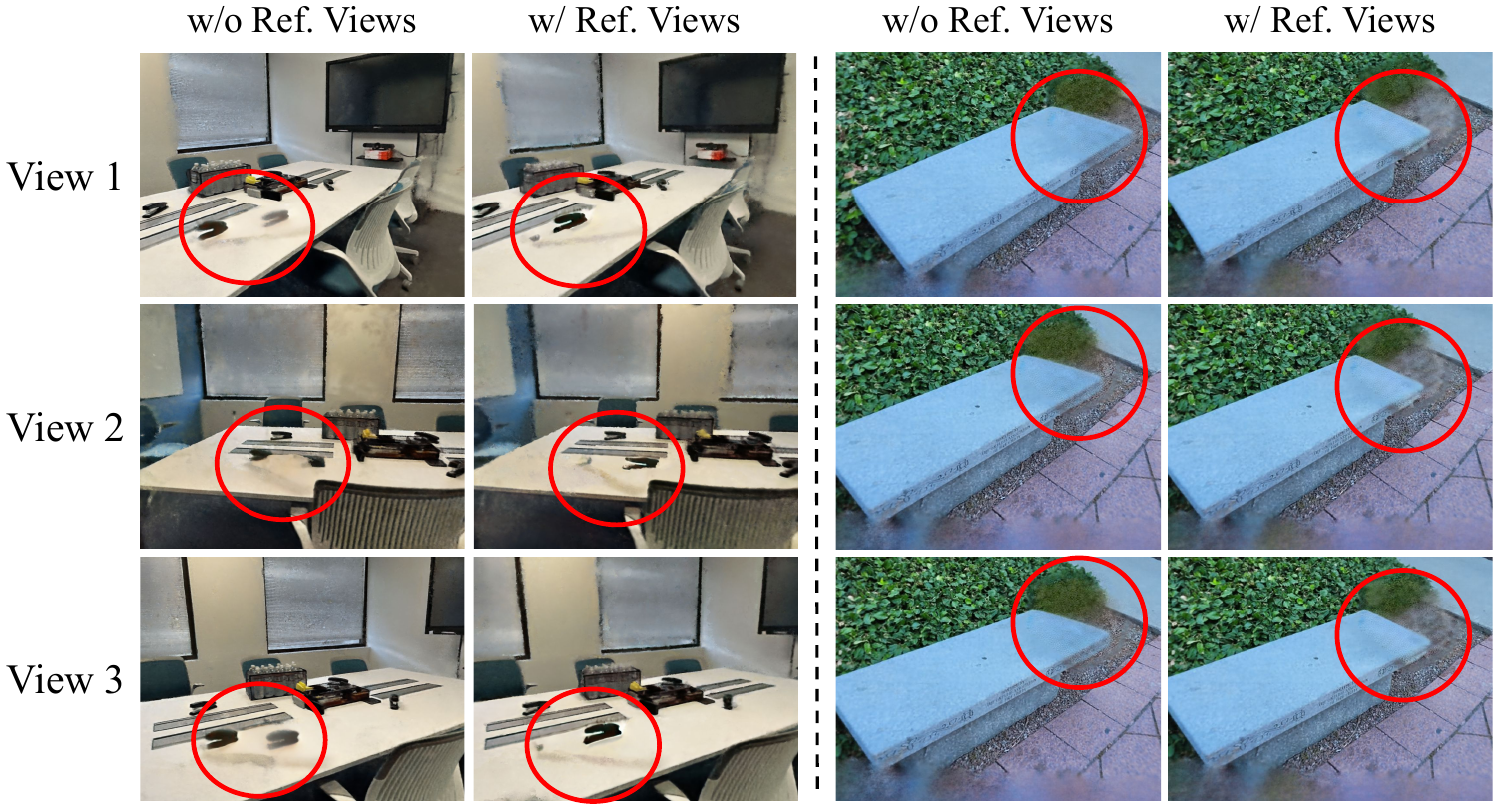}
   \caption{With Reference Views applied in the training process (bottom row), the NeRF learns to render the masked region with more information globally from the data, as compared to incorrect inpainting while rendering a smaller set of training views only (e.g. the hallucination of the stapler in the left scene or the extension of the bench's corner in the right scene).}
   \label{fig:ref}
\end{figure}
\vspace{-.25em}

\subsection{Reference Views Updating}
\label{subsec:ref}
So far, we have introduced two major techniques in our method that address knowledge sharing across multiple views to enhance occluded reconstruction and cross-view consistency.
As shown in ~\cref{eq:cds_gp}, both techniques distill knowledge from multiple rendered views to the NeRF.
To maximize this knowledge distillation across multiple views, we propose a training paradigm with Reference Views.
Specifically, during each iteration, we rendered a set of training views, \(V_{train}\), to backpropagate the loss to train the NeRF and another set of reference views, \(V_{ref}\), the gradient of which will not be calculated, to provide extra information in both CDS and Grid-denoising.
In ~\cref{eq:cds_gp}, the kernel will be calculated within \(V_{train}\), i.e., \(\mathbf{x}^{(i)}, \mathbf{x}^{(j)}\in V_{train}\).
Meanwhile, grids will be formed by images from \(V_{train} \cup V_{ref}\), i.e. \(\{ \mathbf{z}^{(s)}_t \}_{s=1}^S \subset (V_{train} \cup V_{ref})\).

\begin{figure*}[htp]
  \centering
  \includegraphics[trim={1.2cm 0cm 3.5cm 0cm},clip,width=.8\linewidth,page=7]{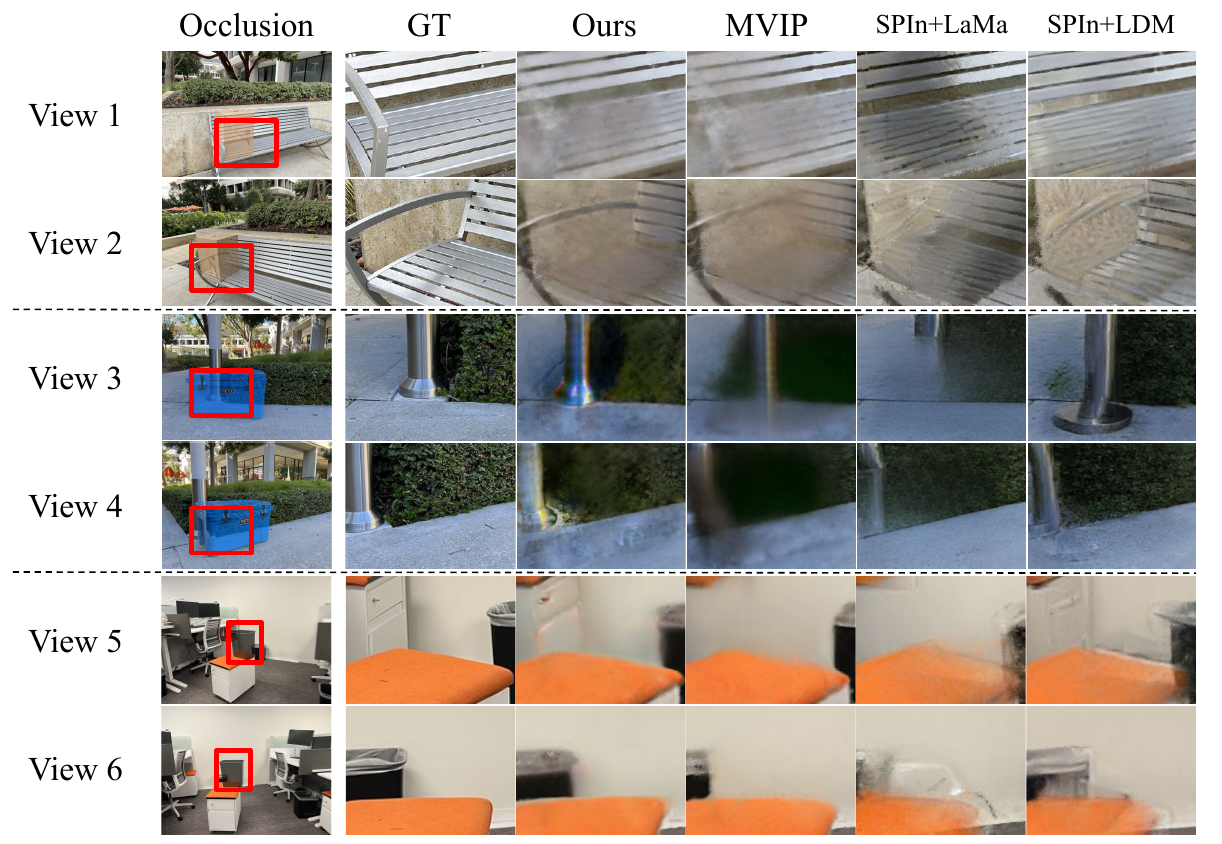}
   \caption{Visualization of qualitative results. For scenes that do not have much occlusion (View 1\&2, the armrest is visible in most views), our method and the baseline methods can render similarly good novel views in terms of quality and fidelity.
   When challenged with severe occlusion (View 3-6), our method excels.
   Specifically, while baseline methods might render correctly in the views where occlusion is not severe, they cannot propagate the information to other views.
   For example, MVIP can correctly render the base of the pole in View 4, but it cannot render a consistent base in View 3.
   During distillation, such inconsistency is propagated and results in artifacts and blurriness in both views.
   Similar examples can be found in View 5\&6 (incorrect positions of the trash bin).
   One limitation is that like prior methods, our method cannot remove the shadow.}
   \label{fig:qualitative}
\end{figure*}

We illustrate how training with Reference Views enhances the consistency and information sharing qualitatively in ~\cref{fig:ref} and quantitatively in our ablation studies.

\subsection{Per-scene Fine-tuning}
\label{subsec:ft}
\begin{table*}[htp]
\centering
\resizebox{\textwidth}{!}{
\begin{tabular}{lcccccccccc}
\toprule
 & \multicolumn{5}{c}{\textbf{SPIn-NeRF}} & \multicolumn{5}{c}{\textbf{Occlude-NeRF}} \\
\cmidrule(lr){2-6} \cmidrule(lr){7-11}
 & PSNR $\uparrow$ & LPIPS $\downarrow$ & L2 $\downarrow$ & SSIM $\uparrow$ & Corrs. $\uparrow$ & PSNR $\uparrow$ & LPIPS $\downarrow$ & L2 $\downarrow$ & SSIM $\uparrow$ & Corrs. $\uparrow$\\
\midrule
SPIn-NeRF+LaMa & 15.992 & \textbf{0.284} & 0.287 & 0.416 & 73.080 & 14.154 & 0.321 & 0.369 & 0.619  & 37.483 \\
SPIn-NeRF+LDM & 16.162 & 0.298 & \textbf{0.286} & 0.408 & 79.640 & 14.107 & \textbf{0.284} & 0.338 & 0.615  & 47.442 \\
MVIP-NeRF & 16.080 & 0.404 & 0.302 & 0.449 & 76.515 & 13.872 & 0.387 & 0.345 & 0.601  & 37.023 \\
Occlude-NeRF & \textbf{16.177} & 0.355 & 0.290 & \textbf{0.461} & \textbf{86.715} & \textbf{15.447} & 0.346 & \textbf{0.314} & \textbf{0.680} & \textbf{50.928} \\
\bottomrule
\end{tabular}
}
\caption{Evaluation results for different methods on the SPIn-NeRF and Occlude-NeRF datasets.}
\label{tab:comparisons}
\vspace{-1em}
\end{table*}

\begin{table*}[h]
\centering
\resizebox{\textwidth}{!}{
\begin{tabular}{lcccccccccc}
\toprule
 & \multicolumn{5}{c}{\textbf{SPIn-NeRF}} & \multicolumn{5}{c}{\textbf{Occlude-NeRF}} \\
\cmidrule(lr){2-6} \cmidrule(lr){7-11}
 & PSNR $\uparrow$ & LPIPS $\downarrow$ & L2 $\downarrow$ & SSIM $\uparrow$ & Corrs. $\uparrow$ & PSNR $\uparrow$ & LPIPS $\downarrow$ & L2 $\downarrow$ & SSIM $\uparrow$ & Corrs. $\uparrow$\\
\midrule
(i) Ours w/o CDS & 14.802 & 0.389 & 0.329 & 0.383 & 85.517 & 14.841 & 0.415 & 0.315 & 0.615 & 45.413  \\
(ii) Ours w/o G.D. & 15.684 & 0.367 & 0.297 & 0.449 & 81.498 & 14.774 & 0.361 & 0.324 & 0.628 & 42.888  \\
(iii) Ours w/o Ref. & 16.037 & 0.372 & 0.318 & 0.453 & 82.996 & 14.867 & 0.356 & 0.327 & 0.620 & 45.698  \\
(iv) Ours w/o F.T. & 15.562 & 0.358 & 0.305 & 0.449 & 83.844 & 13.661 & 0.397 & 0.368 & 0.602 & 46.412 \\
(v) Ours (full) & \textbf{16.177} & \textbf{0.355} & \textbf{0.290} & \textbf{0.461} & \textbf{86.715} & \textbf{15.447} & \textbf{0.346} & \textbf{0.314} & \textbf{0.680} & \textbf{50.928} \\
\bottomrule
\end{tabular}
}
  \caption{Evaluation results for different ablations of our methods on the SPIn-NeRF and Occlude-NeRF datasets.}
  \label{tab:ablations}
\vspace{-1em}
\end{table*}

As pointed out by prior works~\cite{lin2025taming}, applying latent diffusion models to real-world content such as NeRF is usually challenged from converging to a crisp and deterministic geometry and texture, due to the high diversity of synthetic content from diffusion models.
In pursuit of a more realistic and blending-in convergence of the inpainting, we fine-tune the diffusion model for each scene individually.
Specifically, we choose a fixed text token for each scene and apply LoRA~\cite{hu2021lora} to fine-tune the U-Net of the diffusion model.
Each training sample is generated by randomly masking a view from the training views with one rectangle or one circle (\(p=0.5\) each), with the masked region being the ground truth.
We utilize prior-preservation loss~\cite{ruiz2023dreambooth} and MSE to supervise the learning. 
To avoid the model learning patterns from the area we aim to inpaint in NeRF, we mask the training views with the original masks for NeRF training and set the loss within to zero during fine-tuning.

\section{Experiments}
\label{sec:exps}


\subsection{Experiment Setup}
\subsubsection{Baseline Methods}
We chose SPIn-NeRF~\cite{mirzaei2023spin} with LaMa~\cite{rombach2021highresolution}, SPIn-NeRF with Latent Diffusion Model (LDM), and MVIP-NeRF~\cite{chen2024mvip} as our baseline methods for the 3D NeRF inpainting task.
The first two baseline methods are representatives of NeRF inpainting with 2D inpainted images and MVIP-NeRF serves as our baseline for SDS-based NeRF inpainting.

\subsubsection{Datasets}
We chose the SPIn-NeRF dataset and a subset of the LLFF~\cite{mildenhall2019llff} dataset annotated (with masks and pseudo-depth maps) by SPIn-NeRF~\cite{mirzaei2023spin} for our experiments, to assess the algorithms' performance in general cases where occlusion occurs regularly.
Additionally, we collected a novel dataset for NeRF inpainting, named Occlude-NeRF.
This dataset was deliberately crafted with severe occlusion in the scene to provide a rigorous test of the algorithms' performance.
The detailed procedure for creating this dataset is provided in Supplementary~\cref{sec:supp_dataset_building}. 
In the SPIn-NeRF dataset, there are \(10\) scenes each with 60 training views and 40 testing views with their corresponding masks of an object to remove.
The LLFF dataset contains varying numbers of images in \(5\) scenes.
The Occlude-NeRF dataset contains \(6\) scenes, each with 60 training views and 40 testing views.

\subsubsection{Implementation Details}
We implement all baseline methods with hyperparameters reported in their corresponding papers with a few exceptions.
For SPIn-NeRF+LDM, we chose Stable Diffusion 2 (SD 2) Inpainting~\cite{rombach2021highresolution} as the 2D LDM inpainter.
For MVIP-NeRF, we chose SD 2 Inpainting as the diffusion priors, for fair comparison.
In the implementation of our methods, we chose SD 2 Inpainting as the diffusion priors and set the Grid size to \(2\times2\) for computation efficiency.
For the Reference-View training with the SPIn-NeRF and the Occlude-NeRF dataset, we randomly select 12 training views (\(|V_{train}|=12\)) and 48 reference views (\(|V_{ref}=48|\)).
For the LLFF dataset, we set (\(|V_{train}|=4\)) and (\(|V_{ref}=16|\)) due to the smaller sample sizes.
More detailed hyperparameters are listed in Supplementary ~\cref{sec:supp_hyperparameters}. 

\subsection{Results}

\subsubsection{Metrics} 
In our experiments, we aim to assess the following attributes:
\noindent\textbf{Quality and Fidelity}: The visual characteristics and realism of the images.
\noindent\textbf{Faithfulness}: How accurately the rendered image corresponds to the original scene.
\noindent\textbf{Cross-view Consistency}: How the rendered views remain consistent across different viewpoints

We follow similar prior works~\cite{lin2025taming,chen2024mvip,mirzaei2023spin} and evaluate the quality with PSNR and fidelity with LPIPS~\cite{zhang2018perceptual}.
We apply L2 pixel-wise error and SSIM to assess the faithfulness of the synthesized views.
We evaluate the cross-view consistency of the rendered views with a Correspondence Score (Corrs.), similar to~\cite{weber2024nerfiller}, where we report the numbers of high-quality LoFTR~\cite{sun2021loftr} correspondences identified in 100 randomly sampled pairs of rendered images and their ground truths.
To balance the randomness, we proceed for 20 iterations and take the averages.
Note that, unlike some prior work, we do not evaluate with FID~\cite{heusel2017gans} or KID~\cite{binkowski2018demystifying}, because the NeRF datasets are relatively small with insufficient data points for stable calculation for such metrics~\cite{lucic2018gans,barratt2018note,kodali2017convergence}.
We only evaluate qualitatively on the LLFF dataset due to the absence of removal ground truth.

\subsubsection{V.S. Baselines}
We conduct quantitative evaluations to compare the efficacy of our method v.s. that of three baselines in 3D NeRF inpainting tasks.
Specifically, we focus on the quality of the synthesized views and their faithfulness in reconstructing the original scenes.
The results are reported in ~\cref{tab:comparisons}.
Further visual comparisons are illustrated in ~\cref{fig:qualitative}.

From the quantitative results, we observe that our method achieves the highest SSIM and Corrs. on both datasets, demonstrating a stronger ability of structural preservation, and cross-view consistency.
Both SPIn-NeRF methods yield lower LPIPS scores, which can be attributed to the minimization of LPIPS distance between the inpainting and rendering in their methods.
We also see that on the SPIn-NeRF dataset, where there is not much occlusion, our method yields a close but slightly more L2 distance.
While on the Occlude-NeRF dataset, where occlusion is severe in the scenes, our method outperforms all baseline methods in L2, SSIM, and Corrs, which indicates better consistency and faithfulness.
Such results demonstrate that our method excels in tackling severely occluded 3D inpainting tasks while preserving satisfactory performance in image quality and fidelity but compromising a bit of perceptual coherency.
More qualitative discussions can be found in Supplementary ~\cref{sec:supp_qualitative}.

\subsubsection{Ablation Studies}
We conduct ablation studies on our method to investigate the effect of each of our modules.
We test the ablations on the SPIn-NeRF dataset and the Occlude-NeRF dataset.
Specifically, we start with the full method and ablate (i) CDS, (ii) Grid-based Denoising, (iii) Reference Views, and (iv) Per-scene Fine-tuning, respectively.
The quantitative results are reported in ~\cref{tab:ablations}.

Particularly, comparing ours with (i), we identify a significant drop in L2 and SSIM performance without CDS.
Additionally, we see close Corrs. scores on the SPIn-NeRF dataset but a tremendous improvement of Corrs scores on the Occlude-NeRF dataset.
This affirms the efficacy of CDS in faithful reconstructions across the views, especially in occlusion cases.
Showing by the differences between (v) and (iii), the use of Reference views further enhances the efficacy of CDS in faithfulness with slightly improved image quality and fidelity.
A significant drop in Corrs. (ii) denoising demonstrates that Grid-based denoising greatly contributes to the consistency among the views.
As also suggested by the drop in PSNR and LPIPS scores of (iv) on both datasets, we conclude that per-scene fine-tuning enhances the overall quality of the view synthesis.




\section{Conclusion}
\label{sec:conclusion}
We introduce Occlude-NeRF, a novel approach to inpaint a 3D NeRF scene.
Our method tackles the occlusion challenge unaddressed by prior work, where most views are occluded leaving a limited amount of information about the area to inpaint.
Specifically, we propose a multi-view version of CDS, a Grid-based denoising pattern, and a Reference-view training paradigm to enable information sharing among the views.
We also apply per-scene fine-tuning to enhance the rendering quality and fidelity.
To assess Occlude-NeRF's performance in occlusion cases, we construct a novel dataset with challenging occlusion.
The major limitation of our methods lies in the reconstruction of high-frequency regions due to the averaging of multiple views during distillation, which we further showcase in Supplementary ~\cref{sec:supp_high_freq}.
{
    \small
    \bibliographystyle{ieeenat_fullname}
    \bibliography{main}
}
\clearpage
\setcounter{page}{1}
\maketitlesupplementary

\section{Analysis on Multi-view CDS}
\label{sec:supp_cds}
We have proposed the core distillation sampling strategy in our main manuscript, namely ~\Cref{eq:cds_gp}.
In this Supplementary section, we discuss how ~\Cref{eq:cds_gp} helps propagate the information from limited views to the collaborative update of the NeRF parameters.

\subsection{Kernel-Weighted Noise Prediction}

\subsubsection{Noise Prediction}
For each view \( i \), the noise prediction \( \hat{\mathbf{\epsilon}}^{(i)} \) is computed as a kernel-weighted combination of noise predictions from all views:
\begin{align}
\hat{\mathbf{\epsilon}}^{(i)} = \frac{1}{N} \sum_{j=1}^N k(\mathbf{z}^{(i)}, \mathbf{z}^{(j)}) \mathbf{\epsilon}^{(j)}
\end{align}
where:
\begin{itemize}
    \item \( \mathbf{\epsilon}^{(j)} \) is the noise prediction for view \( j \).
    \item \( k(\mathbf{z}^{(i)}, \mathbf{z}^{(j)}) \) is the kernel function, which measures similarity between the latents \( \mathbf{z}^{(i)} \) and \( \mathbf{z}^{(j)} \).
\end{itemize}

The kernel \( k(\mathbf{z}^{(i)}, \mathbf{z}^{(j)}) \) ensures that:
\begin{itemize}
    \item Views \( j \) are from the rendered set as view \(i\). Those from views \(j\) with relevant information (e.g., visible occluded areas) contribute more strongly to the noise prediction \( \hat{\mathbf{\epsilon}}^{(i)} \) of view \( i \).
    \item Occluded view \( i \) to incorporate details from views \( j \) where the occluded area is visible.
\end{itemize}

\subsection{Loss Function and Gradient}

\subsubsection{Loss Function}
The loss for each view \( i \) is defined as:
\begin{align}
L^{(i)} = \hat{\mathbf{\epsilon}}^{(i)} - \mathbf{\epsilon}_{\text{gt}}^{(i)}
\end{align}
where \( \mathbf{\epsilon}_{\text{gt}}^{(i)} \) is the ground-truth noise for view \( i \). Substituting \( \hat{\mathbf{\epsilon}}^{(i)} \), the loss becomes:
\begin{align}
L^{(i)} = \frac{1}{N} \sum_{j=1}^N k(\mathbf{z}^{(i)}, \mathbf{z}^{(j)}) \mathbf{\epsilon}^{(j)} - \mathbf{\epsilon}_{\text{gt}}^{(i)}
\end{align}

\subsubsection{Gradient of the Loss}
The gradient of this loss with respect to the latent \( \mathbf{z}^{(i)} \) is:
\begin{align}
\nabla_{\mathbf{z}^{(i)}} L^{(i)} = \frac{1}{N} \sum_{j=1}^N \nabla_{\mathbf{z}^{(i)}} \left( k(\mathbf{z}^{(i)}, \mathbf{z}^{(j)}) \right) \mathbf{\epsilon}^{(j)}
\end{align}

For a Gaussian RBF kernel with scale \(h\):
\begin{align}
k(\mathbf{z}^{(i)}, \mathbf{z}^{(j)}) = \exp \left( -\frac{1}{h} \| \mathbf{z}^{(i)} - \mathbf{z}^{(j)} \|^2_2 \right)
\end{align}
the gradient is:
\begin{align}
\nabla_{\mathbf{z}^{(i)}} k(\mathbf{z}^{(i)}, \mathbf{z}^{(j)}) = -\frac{2}{h} (\mathbf{z}^{(i)} - \mathbf{z}^{(j)}) k(\mathbf{z}^{(i)}, \mathbf{z}^{(j)})
\end{align}

Substituting this into \( \nabla_{\mathbf{z}^{(i)}} L^{(i)} \), we get:
\begin{align}
\nabla_{\mathbf{z}^{(i)}} L^{(i)} = \frac{1}{N} \sum_{j=1}^N \left( -\frac{2}{h} (\mathbf{z}^{(i)} - \mathbf{z}^{(j)}) k(\mathbf{z}^{(i)}, \mathbf{z}^{(j)}) \right) \mathbf{\epsilon}^{(j)}
\end{align}

This gradient shows how information propagates between views \( i \) and \( j \), with the kernel \( k(\mathbf{z}^{(i)}, \mathbf{z}^{(j)}) \) modulating the strength of interaction.

\subsection{NeRF Parameter Updates}

\subsubsection{Gradient Propagation}
The latent updates are propagated to the NeRF parameters \( \theta \) through backpropagation. The total gradient for \( \theta \) is:
\begin{align}
\nabla_{\theta} L = \sum_{i=1}^N \nabla_{\theta} L^{(i)}
\end{align}

Using the chain rule:
\begin{align}
\nabla_{\theta} L^{(i)} = \nabla_{\mathbf{z}^{(i)}} L^{(i)} \cdot \nabla_{\theta} \mathbf{z}^{(i)}
\end{align}
Substituting \( \nabla_{\mathbf{z}^{(i)}} L^{(i)} \) from above, we get:
\begin{align}
&\nabla_{\theta} L = \sum_{i=1}^N \nabla_{\theta} \mathbf{z}^{(i)} \cdot \notag \\
&\left( \frac{1}{N} \sum_{j=1}^N \left( -\frac{2}{h} (\mathbf{z}^{(i)} - \mathbf{z}^{(j)}) k(\mathbf{z}^{(i)}, \mathbf{z}^{(j)}) \right) \mathbf{\epsilon}^{(j)} \right)
\end{align}

\begin{figure*}[htp]
  \centering
  \includegraphics[trim={0cm 0cm 0cm 0cm},clip,width=\linewidth,page=8]{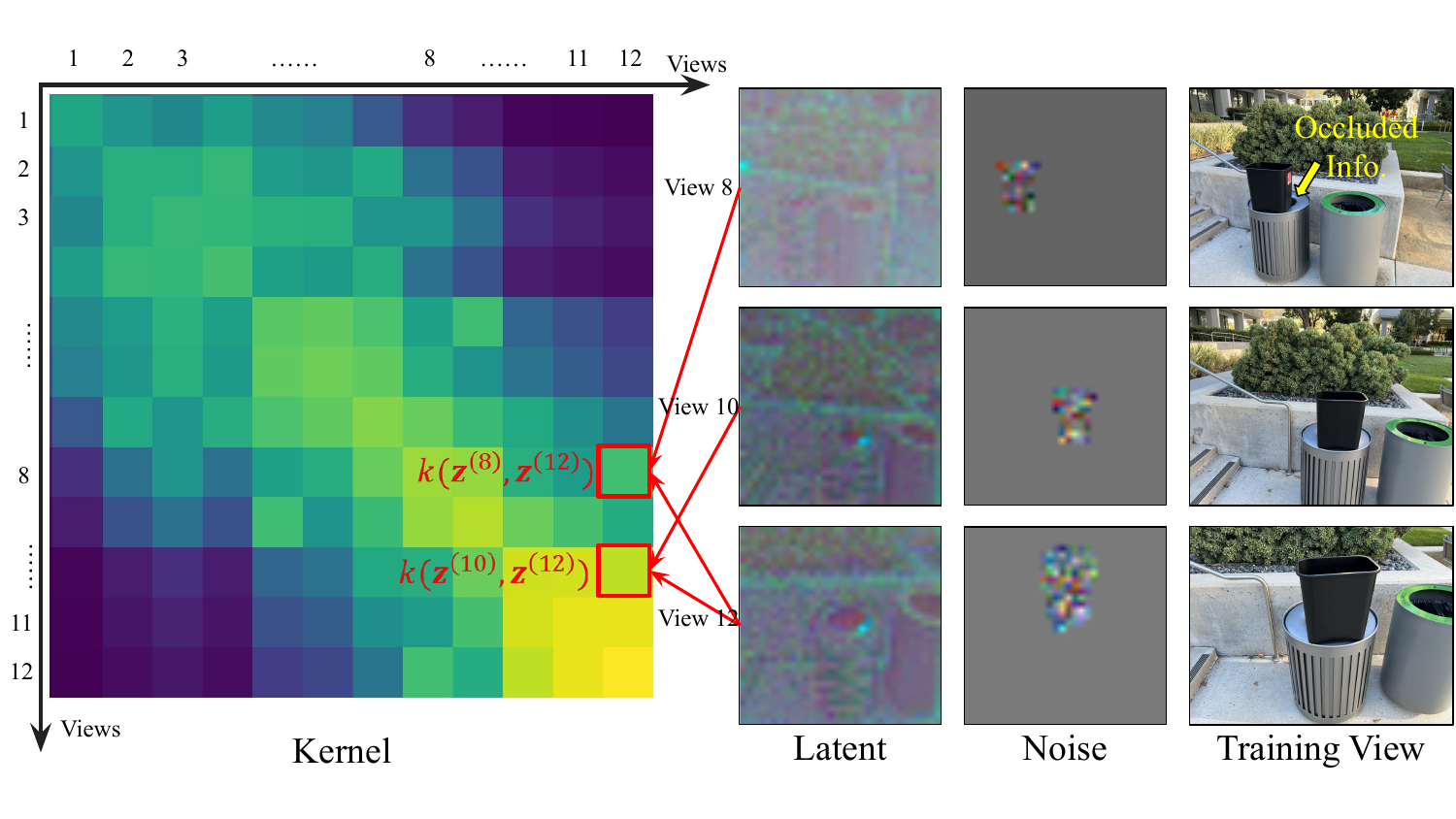}
   \caption{Illustration of the effect of our CDS kernel in one iteration. The heatmap of the kernel is on the left. The warmer the color, the higher the kernel value. We can see the corresponding views on the right-hand side. The closer (in the latent space) the views are, the higher the correspondence value in the kernel (View \(10\) \& \(12\)). Meanwhile, views with further distance but containing important information can also be related to the update (e.g. the kernel has a relatively high value between (View \(8\) \& \(10\)). In this way, the information about the occluded area (e.g. View \(8\), where the hole of the left trash bin is visible) can be propagated to the update of other views (e.g. View \(12\), where the hole is completely occluded).}
   \label{fig:kernel}
\end{figure*}

\subsection{Role of the Two Terms in the Kernel Update}

The kernel update involves two key terms:
\begin{align}
\nabla_{\mathbf{z}^{(i)}} k(\mathbf{z}^{(i)}, \mathbf{z}^{(j)}) = -\frac{2}{h} (\mathbf{z}^{(i)} - \mathbf{z}^{(j)}) k(\mathbf{z}^{(i)}, \mathbf{z}^{(j)})
\end{align}
and:
\begin{align}
k(\mathbf{z}^{(i)}, \mathbf{z}^{(j)}).
\end{align}

\subsubsection*{First Term: Gradient of the Kernel (\( \nabla_{\mathbf{z}^{(i)}} k(\mathbf{z}^{(i)}, \mathbf{z}^{(j)}) \))}

This term serves several critical purposes in the kernel-based updates:

1. \textbf{Repulsive Force to Maintain Diversity}:
   \begin{align}
   \nabla_{\mathbf{z}^{(i)}} k(\mathbf{z}^{(i)}, \mathbf{z}^{(j)}) = -\frac{2}{h} (\mathbf{z}^{(i)} - \mathbf{z}^{(j)}) k(\mathbf{z}^{(i)}, \mathbf{z}^{(j)})
   \end{align}
   The term \( (\mathbf{z}^{(i)} - \mathbf{z}^{(j)}) \) computes the directional vector pointing from \( \mathbf{z}^{(j)} \) to \( \mathbf{z}^{(i)} \). 
   The negative sign ensures that the gradient drives \( \mathbf{z}^{(i)} \) away from \( \mathbf{z}^{(j)} \), creating a repulsive effect between similar latents.
   This repulsion prevents all latents from collapsing into a single representation, ensuring sufficient diversity among the latent representations for different views.

2. \textbf{Propagation of Occlusion Information}:
   When a view \( j \) contains visible information about an occluded area, its latent \( \mathbf{z}^{(j)} \) contributes gradients to the update of \( \mathbf{z}^{(i)} \) through this term:
     \begin{align}
     \nabla_{\mathbf{z}^{(i)}} L^{(i)} = \frac{1}{N} \sum_{j=1}^N \nabla_{\mathbf{z}^{(i)}} k(\mathbf{z}^{(i)}, \mathbf{z}^{(j)}) \mathbf{\epsilon}^{(j)}.
     \end{align}
   If \( \mathbf{z}^{(j)} \) is close to \( \mathbf{z}^{(i)} \), the kernel \( k(\mathbf{z}^{(i)}, \mathbf{z}^{(j)}) \) will be large, amplifying the influence of \( \mathbf{z}^{(j)} \) on \( \mathbf{z}^{(i)} \). This ensures that visible details in \( \mathbf{z}^{(j)} \) are propagated into the occluded representation \( \mathbf{z}^{(i)} \).

3. \textbf{Modulation by Kernel Weight}:
   The term \( k(\mathbf{z}^{(i)}, \mathbf{z}^{(j)}) \) modulates the strength of the gradient, ensuring that only nearby latents significantly influence \( \mathbf{z}^{(i)} \).
   Mathematically, the magnitude of the gradient is proportional to the similarity between \( \mathbf{z}^{(i)} \) and \( \mathbf{z}^{(j)} \), as measured by \( k(\mathbf{z}^{(i)}, \mathbf{z}^{(j)}) \).

\subsubsection*{Second Term: Kernel Weight (\( k(\mathbf{z}^{(i)}, \mathbf{z}^{(j)}) \))}

This term determines how much influence view \( j \) has on view \( i \) in the kernel-weighted noise prediction:
\begin{align}
\hat{\mathbf{\epsilon}}^{(i)} = \frac{1}{N} \sum_{j=1}^N k(\mathbf{z}^{(i)}, \mathbf{z}^{(j)}) \mathbf{\epsilon}^{(j)}.
\end{align}

\begin{figure*}[htp]
  \centering
  \includegraphics[trim={0cm 0cm 0cm 0cm},clip,width=\linewidth,page=9]{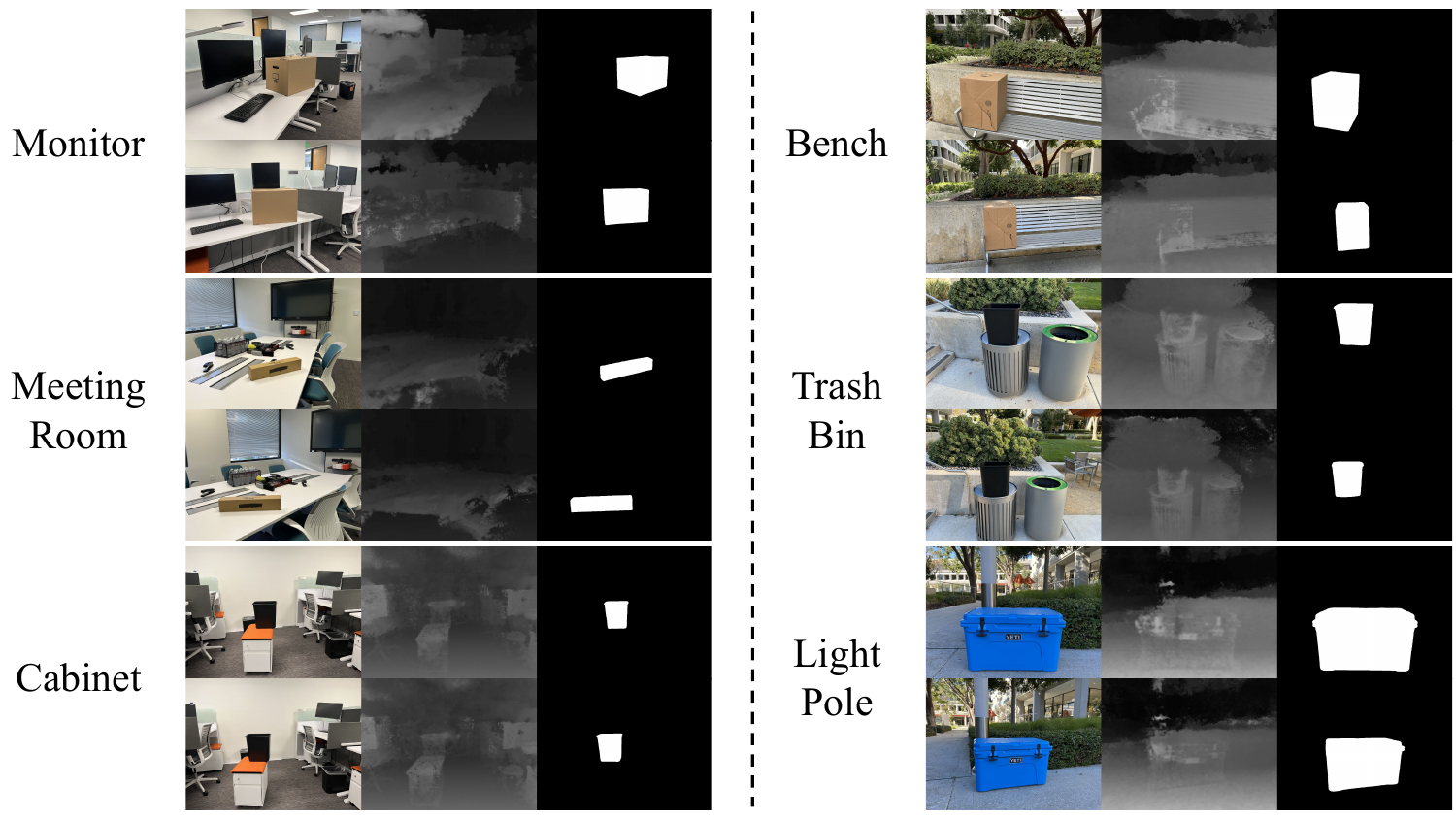}
   \caption{Visualization of our customized dataset. For each scene, we set up an obstacle blocking an area and mask the obstacle for inpainting tasks. We collect testing views and training views with  RGB images and masks. We also generate the pseudo-depth maps (visualized as the disparity maps) corresponding to each view.}
   \label{fig:dataset}
\end{figure*}

1. \textbf{Weighted Contribution of Nearby Views}:
   The kernel \( k(\mathbf{z}^{(i)}, \mathbf{z}^{(j)}) \) assigns higher weights to views \( j \) with similar latents \( \mathbf{z}^{(j)} \) to \( \mathbf{z}^{(i)} \), ensuring that these views have a stronger influence on the noise prediction for view \( i \).
   This is particularly critical when \( \mathbf{z}^{(j)} \) contains visible information about an occluded area in \( \mathbf{z}^{(i)} \), as the kernel amplifies the contribution of \( \mathbf{z}^{(j)} \).

2. \textbf{Occlusion-Aware Updates}:
   For occluded areas, \( \mathbf{z}^{(j)} \) from views where the occlusion is visible dominates the noise prediction for \( \mathbf{z}^{(i)} \), effectively propagating information about the occluded area across views:
     \begin{align}
     \hat{\mathbf{\epsilon}}^{(i)} \approx \frac{\sum^N_{j} k(\mathbf{z}^{(i)}, \mathbf{z}^{(j)}) \mathbf{\epsilon}^{(j)}}{\sum^N_{j} k(\mathbf{z}^{(i)}, \mathbf{z}^{(j)})}.
     \end{align}
   This weighted update ensures that the occluded representation \( \mathbf{z}^{(i)} \) aligns with the visible views.

3. \textbf{Locality of Influence}:
   The kernel decays rapidly with distance in the latent space:
     \begin{align}
     k(\mathbf{z}^{(i)}, \mathbf{z}^{(j)}) = \exp \left( -\frac{1}{h} \| \mathbf{z}^{(i)} - \mathbf{z}^{(j)} \|^2_2 \right).
     \end{align}
   As \( \| \mathbf{z}^{(i)} - \mathbf{z}^{(j)} \| \) increases, \( k(\mathbf{z}^{(i)}, \mathbf{z}^{(j)}) \to 0 \), ensuring that only nearby latents significantly influence the updates.
   In this way, we avoid distillation of 2D prior from views that are too far away, which may result in inconsistent 2D inpainting results, as also pointed out by prior work~\cite{weber2024nerfiller}.

\subsubsection*{Combined Functionality of the Two Terms}

The combined effect of the two terms is as follows:
\begin{itemize}
    \item The \textbf{first term} (\( \nabla_{\mathbf{z}^{(i)}} k \)) ensures that information propagates between views and prevents collapse by introducing repulsive forces.
    \item The \textbf{second term} (\( k \)) amplifies contributions from relevant views, particularly those with visible occluded areas, ensuring effective information sharing.
\end{itemize}

Together, these terms propagate occlusion information across views, align latent representations, and maintain diversity in the latent space, enabling robust NeRF training.

\subsubsection*{Why Occlusion Information Propagates to \( \theta \)}
1. \textbf{Kernel-Based Weighting:}
   The kernel \( k(\mathbf{z}^{(i)}, \mathbf{z}^{(j)}) \) ensures that visible views \( j \) contribute more strongly to occluded views \( i \), propagating occlusion details across the latent space, as shown in ~\Cref{fig:kernel}

2. \textbf{Collaborative Updates to Latents:}
   The gradient \( \nabla_{\mathbf{z}^{(i)}} k(\mathbf{z}^{(i)}, \mathbf{z}^{(j)}) \) drives latent \( \mathbf{z}^{(i)} \) of occluded views to align with \( \mathbf{z}^{(j)} \) of visible views.

3. \textbf{Backpropagation to NeRF:}
   The updated latents \( \mathbf{z}^{(i)} \) are used to refine the NeRF parameters \( \theta \), enabling the model to represent occluded areas consistently across all views.

\section{Dataset Building}
\label{sec:supp_dataset_building}

In this section, we introduce our procedure and details for constructing the Occlude-NeRF dataset.

\subsection{Scene Setup}
We collected data from six scenes in total, with three indoors and three outdoors, as shown in ~\Cref{fig:dataset}.
We name the three indoor scenes: \textit{Cabinet}, \textit{Monitor}, and \textit{Meeting Room}, and the three outdoor scenes: \textit{Bench}, \textit{Trash Bin}, and \textit{Light Pole}.
The specific scene description is listed below:
\begin{itemize}
    \item \textit{Cabinet}: an office workspace with a symmetrical arrangement of two cubicles on either side. Each cubicle includes a white desk, a chair with a gray backrest, and an orange seat cushion. A black trash bin is placed on top of a mobile pedestal with an orange cushion. The black trash bin is masked.
    \item \textit{Monitor}: an office workspace with a white desk and a cardboard box placed in the center. On the desk are two Dell monitors (one visible and turned off), a black keyboard, and a desk lamp on the left. The cardboard box is masked.
    \item \textit{Meeting Room}: an indoor office meeting room with a white conference table surrounded by teal office chairs. On the table are various items, including a long cardboard box, staplers, and a black organizer containing stationery such as pens, highlighters, and sticky notes. The cardboard box is masked.
    \item \textit{Bench}: a cardboard box placed on a silver metal bench in an outdoor area. The bench is positioned next to a concrete planter filled with green shrubs and small rocks. The cardboard box is masked.
    \item \textit{Trash Bin}: two outdoor trash bins in front of a concrete planter with green shrubs. The left bin is metallic with vertical slits, and a black rectangular container is placed on its circular opening. The right bin is smooth, gray, and labeled "Compost" with a green rim. The black container is masked.
    \item \textit{Light Pole}: a bright blue cooler placed on a concrete sidewalk, next to a metal pole and a neatly trimmed hedge. The cooler is masked.
\end{itemize}

\subsection{Data Collection}
For each scene, we collect \(60\) training views and \(40\) testing views.
For each training view, we obtain a mask by prompting a point at the object to mask, using the Segment Anything Model (SAM)~\cite{kirillov2023segany}.
Each mask is dilated with a \(3 \times 3\) kernel for \(3\) iterations.
To obtain relatively accurate camera pose estimations, we mark the objects' location in the scene with a marker, place the object to take one image, and remove the object for another while the camera remains static.
In this way, we obtain a testing view with and without the object in the scene.
We then put the object back according to the mark and move the camera for other views.
We conducted this procedure because we found prior work's ~\cite{mirzaei2023spin} method for estimating camera poses resulted in unstable accuracy since they use COLMAP~\cite{schoenberger2016sfm,schoenberger2016mvs} to perform structure from motion with images with and without objects.
Therefore, in our case, we obtain extra images with objects for the testing views, so that we can estimate the poses for both training views and testing views together.
We then obtain the pseudo depth maps for each view following SPIn-NeRF~\cite{mirzaei2023spin}.
The collected dataset samples can be found in ~\Cref{fig:dataset}.

\begin{figure*}[htp]
  \centering
  \includegraphics[trim={0cm 0cm 0cm 0cm},clip,width=.85\linewidth,page=10]{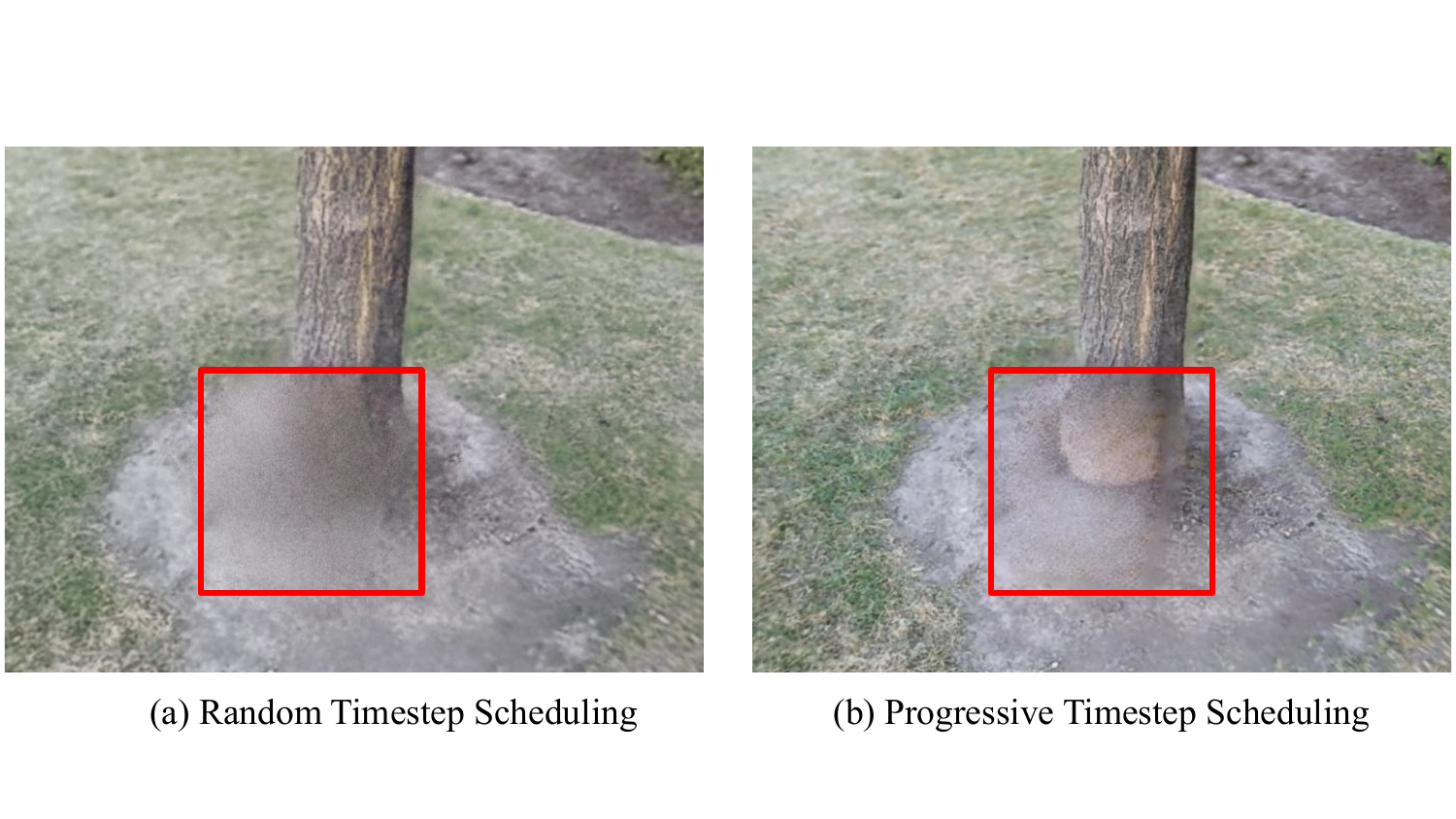}

   \caption{Comparison between (a) random timestep scheduling and (b) progressive timestep scheduling. We can easily observe the better convergence of shape and appearance of the latter.}
   \label{fig:noise_schedule}
\end{figure*}

\section{Hyperparameter Details}
\label{sec:supp_hyperparameters}
In this section, we elucidate the hyperparameters we have used in our experiments, as well as some findings exploring the hyperparameters.
\subsection{Implementation}
We implement our Occlude-NeRF method on 2 NVIDIA H100 GPUs, trained for 10,000 iterations for each scene with the Adam optimizer with a learning rate of \(1e-4\) scheduled with a cosine annealing scheduler (max number of iterations: 50 and min learning rate:0).
For the distillation sampling, we follow prior work~\cite{chen2024mvip,weber2024nerfiller} and choose timesteps uniformly increasing with the training from \(t_{min}=0.02\) to \(t_{max}=0.98\).
For the classifier-free guidance~\cite{ho2022classifier}, we choose a uniform value for all scenes to see the generalizability of our methods, in contrast to MVIP-NeRF~\cite{chen2024mvip}.
Specifically, we set:
\begin{align}
    \hat{\mathbf{\epsilon}} = \mathbf{\epsilon}_{uncond} + \gamma \times \left(\mathbf{\epsilon}_{text}-\mathbf{\epsilon}_{uncond}\right)
\end{align}
where \(\hat{\mathbf{\epsilon}}\) is the final noise prediction, \(\mathbf{\epsilon}_{uncond}\) and \(\mathbf{\epsilon}_{text}\) are the noise prediction with no condition and conditioned by text, respectively, and \(\gamma\) is the guidance scale, which we set uniformly \(\gamma=7.5\).
During training, the size of latent \(z\) is set to \(256\times 256\) for collaborative distillation and \(512\times 512\) for geometry distillation due to GPU RAM limits.
For each iteration, we set the batch (of rays) size to \(1024\) and the number of samples along the ray to \(32\).
Additionally, we set the number of samples for fine networks to \(32\).
We test with three different numbers of the Grid-based Denoising \(M=1,4,8\) and choose \(M=4\) to balance performance and training time.
The textual prompts to diffusion models are listed in ~\Cref{tab:prompts}:

\begin{table*}[h!]
\centering
\renewcommand{\arraystretch}{1.5} 
\setlength{\tabcolsep}{8pt} 

\begin{tabular}{|>{\centering\arraybackslash}l|>{\centering\arraybackslash}l|>{\raggedright\arraybackslash}p{8cm}|}
\hline
\textbf{Dataset Name} & \textbf{Scenes} & \textbf{Prompts} \\ \hline
\multirow{10}{*}{SPIn-NeRF} & 1 & "a stone park bench" \\ \cline{2-3}
                            & 2 & "a wooden tree trunk on dirt" \\ \cline{2-3}
                            & 3 & "a red fence" \\ \cline{2-3}
                            & 4 & "stone stairs" \\ \cline{2-3}
                            & 7 & "a grass ground" \\ \cline{2-3}
                            & 9 & "a corner of a brick wall and a carpeted floor" \\ \cline{2-3}
                            & 10 & "a wooden bench in front of a white fence" \\ \cline{2-3}
                            & 12 & "grass ground" \\ \cline{2-3}
                            & Book & "a brick wall with an iron pipe" \\ \cline{2-3}
                            & Trash & "a brick wall" \\ \hline
\multirow{6}{*}{Occlude-NeRF}  & Monitor & "a computer monitor on a white office desk" \\ \cline{2-3}
                            & Meeting Room & "a black stapler on a white office table" \\ \cline{2-3}
                            & Bench & "a silver metallic bench with slats and armrests" \\ \cline{2-3}
                            & Trash Bin & "a metallic garbage bin with a round opening" \\ \cline{2-3}
                            & Light Pole & "a metallic light pole next to a green ground plant" \\ \cline{2-3}
                            & Cabinet & "a white filing cabinet with an orange cushion on top, next to a white wall" \\ \hline
\multirow{5}{*}{LLFF}  & Fern & "plant and planter on dirt" \\ \cline{2-3}
                            & Fortress & "a wooden tabletop" \\ \cline{2-3}
                            & Horns & "glass windows and a white support pillar on carpet" \\ \cline{2-3}
                            & Orchids & "a conference room with black office chairs and brown carpet" \\ \cline{2-3}
                            & Room & "green leaves of a plant" \\ \hline
\end{tabular}

\caption{Textual prompts used for each scene in our experiments.}
  \label{tab:prompts}
\end{table*}

\subsection{Exploratory Study on Hyperparameters}
In addition to the hyperparameter choices we have reported above, we explore the hyperparameter space and report interesting findings in this subsection.

\subsubsection{Noise Scheduling}
\label{sec:supp_noise_schedule}
We test two noise scheduling methods.
Namely, we first implemented a random sampling schedule, where a random noise timestep between \(t_{min}\) and \(t_{max}\) is chosen.
We then implemented a progressive sampling schedule similar to ~\cite{chen2024mvip, zhu2023hifa}:
\begin{align}
    t = t_{max} - (t_{max}-t_{min})*iter/max\_iter)
\end{align}
where \(iter\) is the current iteration number and \(max\_iter\) is the total number of iterations.

Qualitatively, we found that the progressive sampling schedule fosters convergence toward clearer and sharper inpainting, as shown in ~\Cref{fig:noise_schedule}.
This can be attributed to the larger changes in the earlier stages to form the 3D representations and smaller changes in the later stages, instead of randomly changing the update scale mid-training, which aligns with the similar findings in ~\cite{weber2024nerfiller}.

\subsubsection{Randomization during Grid-based Denoising}
\label{sec:supp_randomization}
\begin{figure}[htp]
  \centering
  \includegraphics[trim={6cm 0cm 6cm 0cm},clip,width=\linewidth,page=11]{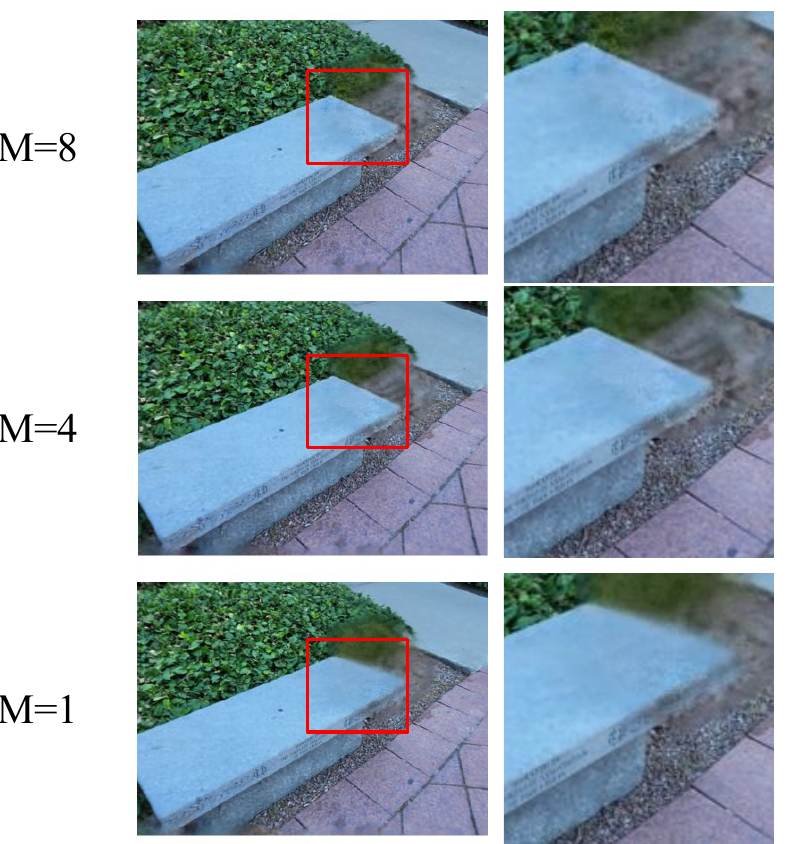}

   \caption{Comparison among different values of \(M\). While \(M=8\) yields the best visual results regarding the sharpness of the inpainting. We made a trade-off between the computation cost in the training phases and the performance and eventually chose \(M=4\) for our experiment.}
   \label{fig:random_M}
\end{figure}

We experimented with different numbers of times shuffling during Grid-based Denoising, namely \(M=1,4,8\).
Our experiments qualitatively showed that \(8\) times of shuffling yields fewer artifacts in the inpainted area, followed by \(M=4\), and then \(M=1\), as shown in ~\Cref{fig:random_M}.
This can be attributed to the averaging effect of the shuffling step in our pipeline, where the influence of multiple views is merged into one update of distillation.
However, increasing \(M\) by one means calling the U-Net for one iteration of denoising, which severely increases the training time.
Therefore, we made a trade-off between training efficiency and performance by setting \(M=4\).

\begin{figure*}[htp]
  \centering
  \includegraphics[trim={0cm 0cm 0cm 0cm},clip,width=.8\linewidth,page=13]{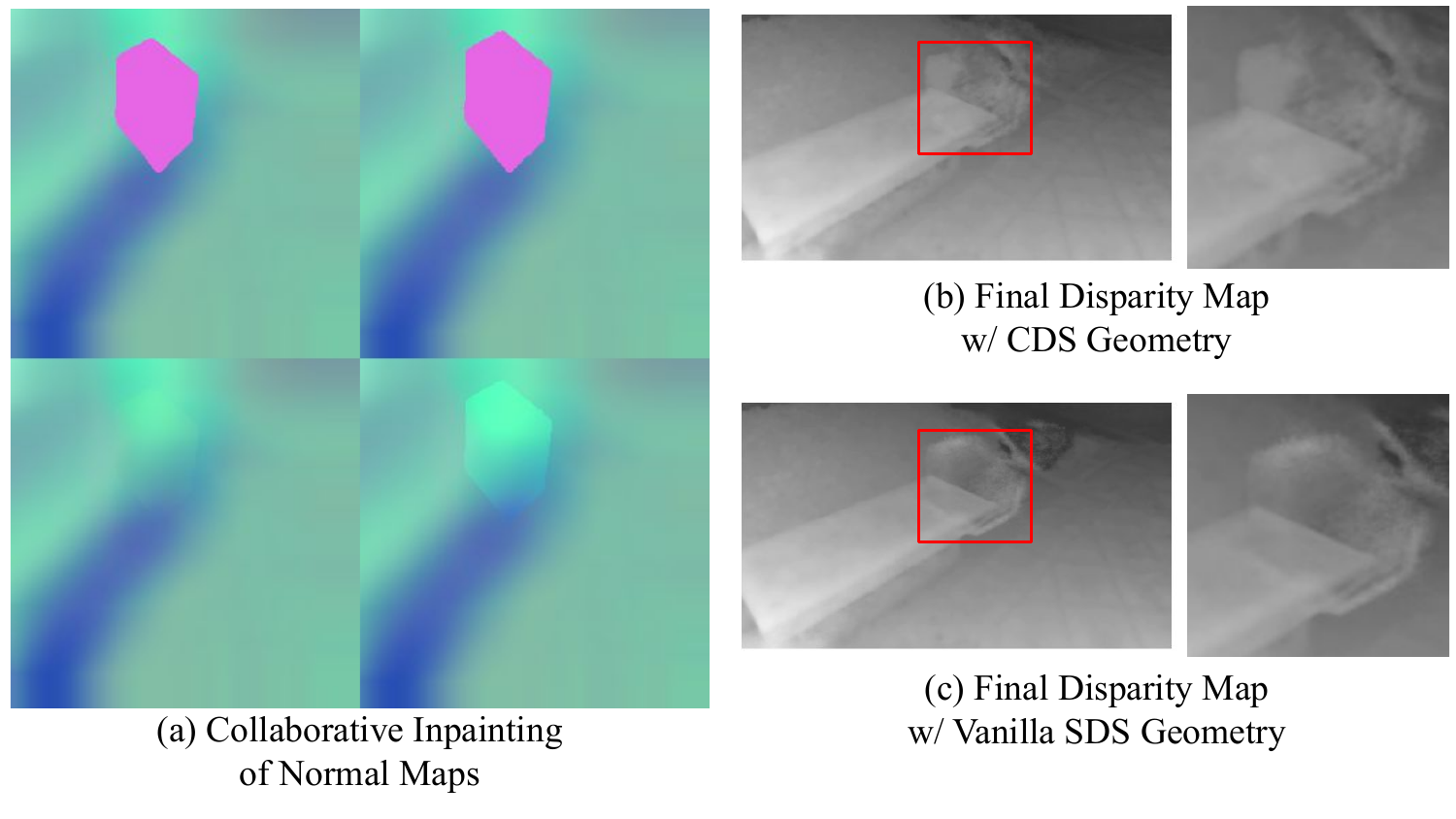}

   \caption{ (a) Collaborative Inpainting with Grid-based denoising does not yield consistent inpainting. The top row is the masked normal maps and the bottom row is the inpainted normal maps (normal maps of the bench scene in SPIn-NeRF). We observed inconsistent inpainting results. Comparing the final results with Vanilla SDS geometry guidance (c) and CDS geometry guidance (b), we observed increased artifacts in the latter.}
   \label{fig:geometry_cds}
\end{figure*}

\subsubsection{Qualitative Comparison with 3D-attention-based SDS}
\begin{figure}[htp]
  \centering
  \includegraphics[trim={4.5cm 0cm 6cm 0cm},clip,width=\linewidth,page=12]{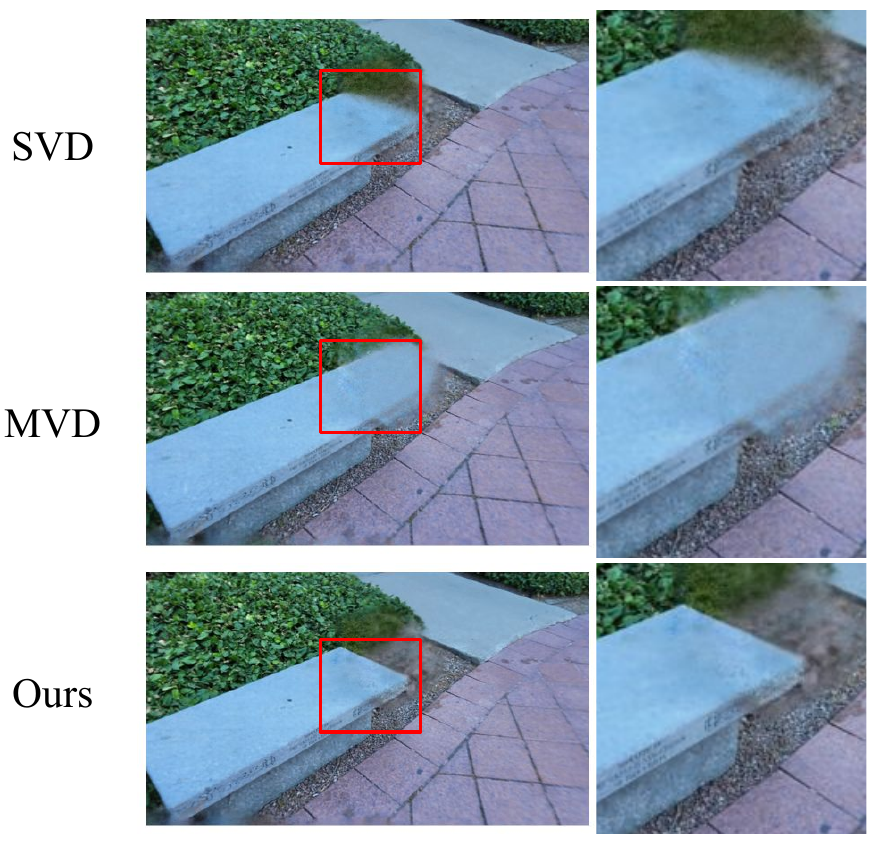}

   \caption{Qualitative comparison between 3D-attention-based diffusion models (SVD \& MVD) and our choice of SD. We observe that SVD does not converge to a sharp reconstruction. Meanwhile, MVD does not yield a correct or faithful reconstruction of the scene.}
   \label{fig:3d_attention}
\end{figure}

Inspired by prior works on 3D-attention in Diffusion Models~\cite{gao2024cat3d, blattmann2023stable, shi2023mvdream}, we attempted 3D-attention in SDS in our early trials.
Specifically, we replaced the diffusion model we used with Stable Video Diffusion (SVD)~\cite{blattmann2023stable} and MVDream (MVD)~\cite{shi2023mvdream}.
For SVD, we trained the NeRF with SD2 for \(3000\) iterations to get the initial reconstruction of the scene and then switched to SVD.
During each iteration, we rendered 14 views with the first one used as the reference and passed then 14 views as a batch to SVD and backpropagated the distillation loss.
For MVD, we disabled the mask condition and passed four rendered views as well as their camera poses as the conditions for the MVD model.
For both methods, the gradients are masked and only enabled in the masked area.
Derived from text-to-image Stable Diffusion~\cite{rombach2021highresolution}, SVD takes a sequence of images as input and allows an additional channel in the input to the U-Net by adding 3D attention layers across the time dimension to compute the self-attention within the batch of input images.
Similarly, MVD utilizes 3D spatial attention to assess the view consistency of the 3D generation from 2D.
The purpose of our trial is to investigate the possibility of directly using such models off-the-shelf in our 3D inpainting tasks to address cross-view consistency.
The visual results are shown in ~\Cref{fig:3d_attention}.
We found that, although efficient in 3D generation tasks, these models are not satisfactory without being further modified and fine-tuned for 3d inpainting tasks, because they do not enable inpainting conditions where the masks and masked images are passed to the U-Net to specify the area and the context to inpaint.
As a result, the 3D generation diffusion models will be prompted to generate 2D prior, which will change the entire image/view rather than just the masked region.
The resulting distillation results do not constitute to a promising 3D inpainting even if we constrain the gradient flow to only the masked region.

\subsection{Vanilla Geometry SDS v.s. CDS Geometry SDS}
\label{sec:supp_geocds}
As a major contribution of our method, we apply collaborative SDS in the color space to tackle the occlusion problem in 3D inpainting.
Yet, for the geometry SDS of a NeRF scene, we only apply vanilla geometry SDS, where we denoise a single normal map per iteration without collaboratively computing the cross-view loss.
This is because, during the experiments, we found that CDS Geometry SDS does not yield consistent distillation in the geometry space.
The inconsistency among the different views of normal maps easily leads to convergence into inpainting with artifacts in the geometry space as shown in ~\Cref{fig:geometry_cds}.

As a result, we do not apply collaborative SDS to the geometry space of our inpainting method.
This can be attributed to the priors in diffusion models being trained mostly on large-scale datasets of natural RGB images paired with textual descriptions (e.g., "a mountain at sunset"), while normal maps are specialized data representing surface orientations using encoded RGB values (usually indicating x, y, z surface normals), which differ fundamentally from natural RGB images in structure and meaning.

\begin{figure*}[htp]
  \centering
  \includegraphics[trim={2.5cm 0cm 2.5cm 0cm},clip,width=\linewidth,page=14]{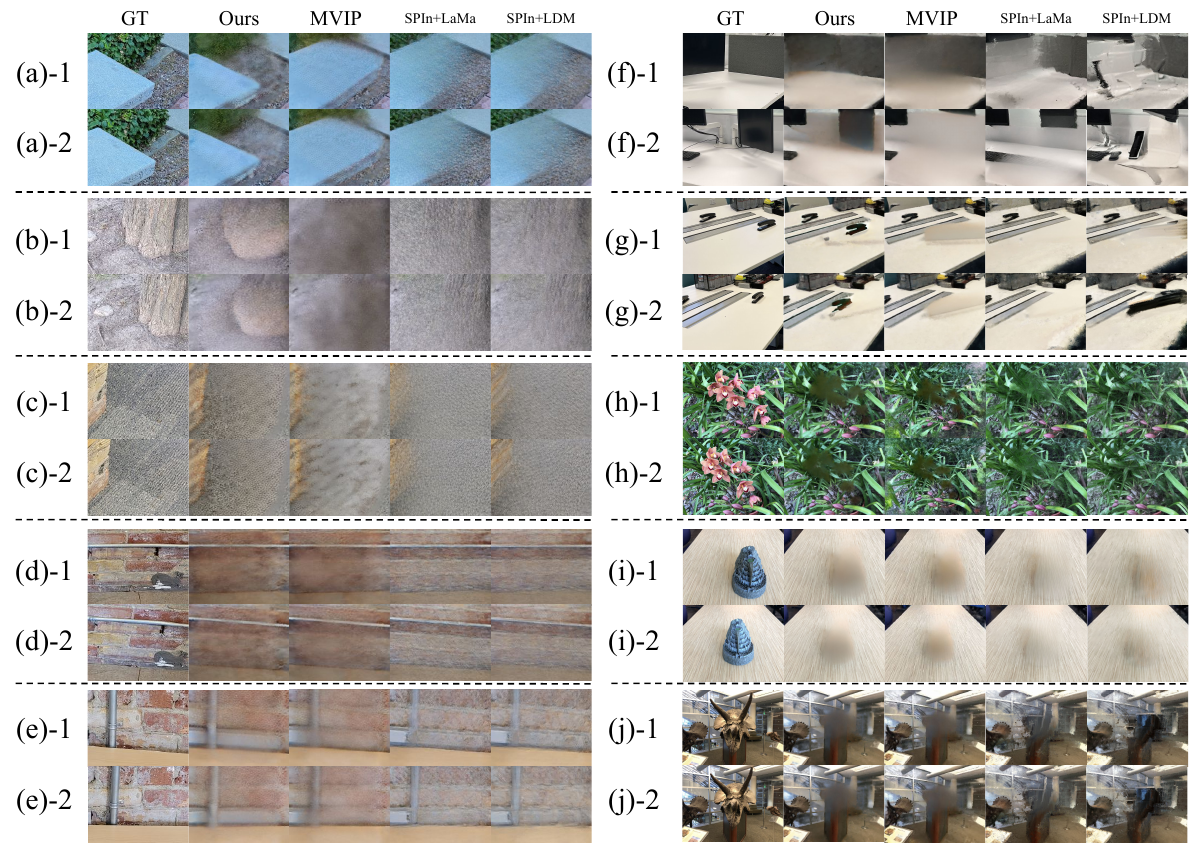}

   \caption{More qualitative results from our experiments. We visualize only the masked region for comparisons. Note that for the LLFF dataset ((h), (i), and (j)), there is no ground truth with the object removed. Thus we show the ground truths with objects instead.
   We observed more faithful reconstructions of the true scenes with our method in severely occluded cases ((a), (b), (c), (f), (g)).
   We also witness more cross-view consistency of our method compared to SPIn-NeRF-based methods ((f) and (j)). However, our method is limited in reconstructing high-frequency texture regions in the scene ((d) and (e) the brick wall, (h) the orchid leaves, and (i) the table), which is a common challenge in SDS-based methods.}
   \label{fig:supp_qualitative}
\end{figure*}

\section{Qualitative Results}
\label{sec:supp_qualitative}
In this section, we present more qualitative results and discuss the shortcomings of our Occlude-NeRF method.

As shown in ~\Cref{fig:supp_qualitative}, our methods can faithfully reconstruct the occluded areas with limited information compared to the baseline methods ( (a) the true edge of the stone bench, (b) the root location of the trunk, (c) the shape of the wall corner, (f) the edge of the monitor, and (g) the location of the stapler).
While being able to reconstruct the occluded area, our method maintains satisfactory visual reconstruction in the cases where occlusion is not severe ((d) and (e) the location of the pipe).

\subsection{Limitation: High-frequency Region Reconstruction}
\label{sec:supp_high_freq}
As prior works~\cite{weber2024nerfiller, lin2025taming} pointed out, recovering high-frequency regions remains a common challenge for 3D generative methods like SDS.
In this subsection, we showcase Occlude-NeRF's limitation in generating high-frequency regions in ~\Cref{fig:supp_qualitative}.
Specifically, in scenes (d) and (e), we observe that the brick wall texture in ours and MVIP's is blurred compared to that in SPIn-NeRF-based methods.
Similarly, in (c) the gray carpet is rendered with artifacts with dot texture instead of the real texture of the carpet.
Moreover, the patterns of the orchid leaves in (h) and the texture of the table in (i) are both blurred in ours and MVIP's, while being sharper and clearer in those in SPIn-NeRF-based methods.
This is attributed to the mechanism of SDS-based methods.
Repetitive updates in SDS will average out the shape of high-frequency objects in the scene.
To tackle this problem, we anticipate future work introducing a more advanced noise scheduling mechanism, where the shape of the reconstruction can be affirmed in the early stages and sharpened in the later stages to avoid blurriness.



\section{Ethical Concerns}
\label{sec:ethics}
The ethical concerns of our method primarily revolve around its potential misuse and implications for privacy, authenticity, and societal impact~\cite{shi2023hci,shi2023understanding}.
Similar algorithms have been applied to editing humanoid avatars~\cite{ma2024avattar, shi2025caring} and objects~\cite{jain2023ubi,he2023ubi} in virtual environments.
The capability of this type of algorithm might be exploited to fabricate or manipulate digital evidence, misrepresent physical spaces, or breach privacy by reconstructing obscured or private areas without consent.
Additionally, biases inherent in training datasets could lead to unfair or inaccurate reconstructions, potentially reinforcing stereotypes or producing misleading results, this is an especially common downfall of text-prompted generative AIs.



\end{document}